\documentclass[12pt]{iopart}

\usepackage[utf8]{inputenc}
\usepackage{a4wide}

\expandafter\let\csname equation*\endcsname\relax

\expandafter\let\csname endequation*\endcsname\relax

\usepackage{amsmath}
\usepackage{amssymb,amsthm,mathrsfs}
\usepackage{dsfont}
\usepackage{graphicx}
\usepackage{tensor}
\usepackage{color}

\usepackage{verbatim}
\usepackage{epstopdf}
\usepackage{xcolor}

\newcommand{\bra}[1]{\langle#1|}
\newcommand{\ket}[1]{|#1\rangle}
\newcommand{\braket}[2]{\langle#1|#2\rangle}
\newcommand{\dyad}[1]{|#1 \rangle\langle #1|}
\newcommand{\dyadB}[2]{|#1 \rangle\langle #2|}
\newcommand{\ketbra}[2]{|#1\rangle\langle #2|}
\newcommand{\braketA}[3]{\langle#1|#2|#3\rangle}

\newcommand{\abs}[1]{\left|#1\right|}

\begin{document}

\title{Two-particle Hadamard walk on dynamically percolated line and circle}

\author{M. Parýzková, M. \v{S}tefa\v{n}\'{a}k$^*$, J. Novotný}
\address{Department of Physics, Faculty of Nuclear Sciences and Physical Engineering, Czech Technical University in Prague, B\v rehov\'a 7, 115 19 Praha 1 - Star\'e M\v esto, Czech Republic}

\author{B. Kollár and T. Kiss}
\address{Wigner Research Center for Physics, SZFKI, Konkoly-Thege M. u. 29-33, H-1121 Budapest, Hungary \\
$^*$ corresponding author }

\ead{martin.stefanak@fjfi.cvut.cz}
\vspace{10pt}
\date{\today}

\begin{abstract}
Asymptotic dynamics of a Hadamard walk of two non-interacting quantum particles on a dynamically percolated finite line or a circle is investigated. We construct a basis of the attractor space of the corresponding random-unitary dynamics and prove the completeness of our solution. In comparison to the one-particle case, the structure of the attractor space is much more complex, resulting in intriguing asymptotic dynamics. General results are illustrated on two examples. First, for circles of length not divisible by 4 the boundary conditions reduces the number of attractors considerably, allowing for fully analytic solution. Second, we investigate line of length 4 and determine the asymptotic cycle of reduced coin states and position distributions, focusing on the correlations between the two particles. Our results show that a random unitary evolution, which is a combination of quantum dynamics and a classical stochasticity, leads to correlations between initially uncorrelated particles. This is not possible for purely unitary evolution of non-interacting quantum particles. The shared dynamically percolated graph can thus be considered as a weak form of interaction. 
\end{abstract}

\maketitle

\section{Introduction}

Quantum walks \cite{Aharonov1993,Meyer1996,Farhi1998} emerged in the 1990's as extensions of classical random walks to a unitary evolution of a quantum particle on a graph or a lattice. Additional twist emerges when we consider an open quantum dynamics \cite{breuer_theory_2007,bengtsson_geometry_2008,alicki_quantum_2014}. A prominent example of this approach are open quantum random walks \cite{attal_open_2012,sinayskiy_open_2013}, which are quantum walks driven by interaction with the external environment. Another possibility is to consider quantum walks in media randomized by static or dynamical percolation \cite{Percolation2}. For discrete-time quantum walks on infinite lattices dynamical percolation was considered as a source of decoherence \cite{romanelli_decoherence_2005,oliveira_decoherence_2006}, and the research focused on spreading properties and transition to classical diffusive behaviour \cite{Abal2008,annabestani_decoherence_2010,Kendon2010,Romanelli2011}, non-markovianity of the evolution of the reduced coin state \cite{hinarejos_chirality_2014}, or mixing times \cite{ghosal_quantum_2018}. On finite graphs, the effect of percolation on quantum transport was investigated \cite{stefanak_lazy,yalcinkaya_coherent_2017,mares2019,Mares_2020_a,Mares_2020_b,Mares_22}. Discrete time percolated quantum walk on a graph with three vertices was experimentally realized \cite{elster2015} using photonic time multiplexing platform \cite{Nitsche_2016}. The influence of static percolation on the efficiency of the discrete-time quantum walk search was studied in \cite{lovett_quantum_2019}. The effect of static percolation of the square lattice on the spreading properties of a discrete quantum walk was recently investigated in \cite{duda_quantum_2023}. In the staggered model of the discrete time quantum walks percolation was introduced recently \cite{santos_decoherence_2022}. In the continuous time case, the influence of static percolation \cite{Mulken2010,Mulken2011,Anishchenko2012,anishchenko_geometrical_2013} and dynamical percolation \cite{Darazs2013,rossi_continuous-time_2017,benedetti_continuous-time_2018} was also thoroughly investigated. 

Asymptotic evolution of the discrete time percolated quantum walks on finite graphs is well understood \cite{Kollar2012}. The investigation utilizes the attractor space of corresponding random unitary operation \cite{Novotny2009,Novotny2010}. The specific form of the quantum walk evolution operator allows to split the equations for a general attractor into the so called coin and shift conditions, which have to be satisfied independently. This simplifies the determination of attractors considerably. In fact, majority of the asymptotic space consists of so called p-attractors, which are constructed from joint eigenstates of evolution operators for all configurations of the percolated graph. Utilizing this method, the asymptotic evolution of percolated quantum walks on a line with a general $SU(2)$ coin  was determined \cite{One_dim_line_general_coin}. The analysis was later extended to 2D quantum walks \cite{Kollar2014}.  

The results discussed above concern a single particle walk. Less is known about the multiparticle case. Two-particle quantum walk on a line with dynamical or static percolation was numerically investigated in \cite{Two_percolations_no_interaction,Two_percolations_interaction}. In the present paper we elaborate in this direction of research. We focus on the asymptotic dynamics of the two-particle dynamically percolated quantum walk on a finite line or a circle. Employing the earlier results for the single particle case \cite{One_dim_line_general_coin} and the particle exchange symmetry, we build a set of linearly independent attractors. In addition, we find genuine two-particle attractors which are not separable. Moreover, we prove the completeness of our solution, i.e. the constructed set of attractors can be turned into an orthonormal basis of the attractor space after orthogonalization procedure. In comparison with the single particle scenario the dimension of the attractor space increases considerably, allowing for much richer asymptotic dynamics.    

The rest of the paper is organized as follows. In Section~\ref{sec:pqw} we review the basic ideas of a single particle percolated quantum walk, including the asymptotic evolution which is determined by the attractors of the corresponding random unitary operation. Section~\ref{sec:2pqw} extends the concepts to two-particle quantum walk. In Section~\ref{sec:construction} we provide a construction of the basis of the attractor space. Technical details, involving the explicit forms of the coin and shift conditions, and the proof of completeness of the constructed basis, are left for appendices. Section~\ref{sec:not4} and \ref{sec:length4} are dedicated to the examples. We conclude and present an outlook in Section~\ref{sec:concl}. 

\section{Percolated quantum walks}
\label{sec:pqw}

Discrete-time quantum walk of one particle in one dimension takes place on the Hilbert space 
\begin{equation}\label{Hilbert space line}
	\mathcal{H} = \mathcal{H}_p \otimes \mathcal{H}_c,
\end{equation}
where $\mathcal{H}_p$ is called the position and $\mathcal{H}_c$ is called the coin space. For a finite line or a circle of length $N$ these are defined through following orthonormal bases
\begin{equation}
	\mathcal{H}_p = \left\lbrace \ket{m}|\, m\in\lbrace 0,1,...,N-1\rbrace \right\rbrace ,\quad \mathcal{H}_c = \left\lbrace \ket{L},\ket{R} \right\rbrace.
\end{equation}
Moving on to the evolution of the system, the evolution operator is defined as a unitary operator on $\mathcal{H}$ of the form
\begin{equation}\label{evolution operator}
	U = S C,
\end{equation}
where $S$ is called the shift operator and $C$ called the coin operator. The coin operator represents the 'coin toss' of the quantum walker and is defined as an arbitrary unitary operator acting solely on the coin degrees of freedom. In cases when the same local coin operation $ \mathcal{C} $ is applied on every position the global coin operator can be written down as
\begin{equation}
	C = I \otimes \mathcal{C}.
\end{equation}
This work is focused on the Hadamard walk, which means that the local coin operator is chosen as the Hadamard coin $H$ on every site
\begin{equation}\label{Hadamard}
	\mathcal{C}=H = \frac{1}{\sqrt{2}}\left(\begin{matrix}
		1&1\\
		1&-1
	\end{matrix}\right).
\end{equation}
The shift operator $ S $ is then a unitary operator which represents the jump of the walker to the next vertex according to the current state of the coin. 
Each step of the walk the whole evolution operator is applied, so for the initial state $\ket{\psi(0)}\in\mathcal{H}$ of the walker the state after $t$ steps will be
\begin{equation}
	\ket{\psi(t)} = U^t \ket{\psi(0)}.
\end{equation}

Turning to the dynamically percolated graph, it is a graph whose edges can randomly break and reappear again during the time evolution of the system. Focusing on just one edge $(v_1, v_2)\in E$ of some general graph $G(E,V)$, there is an assigned probability $p_{12}$ with which the edge will be considered broken, i.e. $(v_1, v_2)\notin K$, where $K\subset E$ is the new configuration of edges, in the current step of the evolution. The set of vertices stays the same during the evolution, there are sometimes just no edges connected to the vertex. There are such probabilities of the edge being broken for all edges of the original graph $G(E,V)$. For the quantum walker it means that every step of the walk can take place on a different subgraph of $G(E,V)$, i.e. on different configuration of edges. The probability of a certain configuration $K$ appearing is then given as
\begin{equation}
	p_K = \prod_{(v_i,v_j)\in K}\prod_{(v_k,v_l)\notin K} (1-p_{ij})p_{kl},
\end{equation} 
where $p_{ab}, \forall v_a,v_b \in V$ is the probability of the edge $(v_a,v_b)$ being broken. We assume that $0<p_{ab}<1$ for all edges of the original graph.

The possibility of different configurations of edges results in application of different associated unitary evolution operators $U_K$. In every step of the evolution one of these unitary operations is randomly applied according to the probability distribution $p_K$. As mentioned earlier, for quantum walks the evolution operator consists of the shift operator $S$ and the coin operator $C$. The Hilbert space of the walker stays the same, but it is necessary to take into account the broken edges in the overall dynamics. This is done by changing the shift operator. More specifically, the shift operator is modified, so that it no longer allows the walker to move along the broken edges.

For a walker on a finite line the non-percolated continuing shift operator $S$ has the following form
\begin{align}\label{shift_operator_finite_normal}
	S = &\sum_{m=0}^{N-2}\dyadB{m+1,R}{m,R} + \mathcal{R}\dyadB{N-1,R}{N-1,R} \nonumber \\ +&\sum_{m=1}^{N-1}\dyadB{m-1,L}{m,L} +\mathcal{R}\dyadB{0,L}{0,L},	
\end{align}
where the end of the line is treated as a broken edge and the local reflection operator $\mathcal{R}$
\begin{equation}\label{reflexe_special_local}
	\mathcal{R}=\left( \begin{matrix}
		0 & 1 \\
		1 & 0
	\end{matrix}\right)=\sigma_x ,
\end{equation} 
is applied. For the circle the non-percolated shift operator is
\begin{align}\label{shift-op-circle_normal}
	S = &\sum_{m=0}^{N-2}\dyadB{m+1,R}{m,R} + \dyadB{0,R}{N-1,R}\nonumber\\+&\sum_{m=1}^{N-1}\dyadB{m-1,L}{m,L} +\dyadB{N-1,L}{0,L},	
\end{align}
where periodic boundary conditions have been applied to both ends. The modified shift operator for a given configuration $K$ is then defined as
\begin{align}\label{shift_operator_K}
	S_K = \sum_{(i,i+1)\in K} \dyadB{i+1,R}{i,R}+&\sum_{(i-1,i)\in K}\dyadB{i-1,L}{i,L} + \nonumber\\ +&\sum_{(i,i+1)\notin K}\mathcal{R} \dyad{i,R}+ \sum_{(i-1,i)\notin K} \mathcal{R} \dyad{i,L}.
\end{align}
This means that if there is no edge, the walker stays in place while changing the coin state due to the local reflection operator.

The analysis of quantum walks on percolated graphs follows from the theory of general random unitary operations, which represent a special type of open quantum evolution \cite{breuer_theory_2007,bengtsson_geometry_2008,alicki_quantum_2014}. Random unitary operation $\Phi$ \cite{holevo_statistical_2001} is generally defined as a completely positive trace-preserving map which can be decomposed into
\begin{equation}\label{superoperator}
	\Phi(\rho) = \sum_{i} p_i U_i \rho U_i^{\dagger}.
\end{equation}
Here $\lbrace U_i \rbrace_i$ is a set of unitary operations acting on a given Hilbert space of the system $\mathcal{H}$, $\rho$ is a density matrix representing the state of the quantum system and $p_i$ is a probability distribution such that $p_i>0, \, \forall i\, $ and $ \sum_{i}p_i = 1$. The superoperator $\Phi$ acts on the space $\mathcal{B}(\mathcal{H})$, which is a Hilbert space of all linear operators acting on $\mathcal{H}$. The evolution of the quantum system is then defined as an iterative application of the superoperator $\Phi$ on the initial state of the system $\rho(0)$, so after $t\in\mathbb{N}$ steps of the evolution the state of the system will be
\begin{equation}
	\rho(t) = \Phi^t(\rho(0)) = \Phi(\Phi(.....\Phi(\rho(0)))).
\end{equation}
Generally $\Phi$ is neither hermitian nor normal, however, it can be shown that in the limit of large enough applications of the superoperator $\Phi$ the resulting state will be from the attractor space of the superoperator defined as
\begin{equation}\label{def_atr_space}
	\text{Atr}(\Phi) = \bigoplus_{\lambda  \in \sigma_{\abs{1}}} \text{Ker}(\Phi - \lambda I),
\end{equation}
where $\sigma_{\abs{1}}$ denotes the set of eigenvalues of $\Phi$ with modulus 1. Based on this definition, it can also be shown \cite{Novotny2010} that 
\begin{equation}\label{konstrukce_attraktoru}
	\text{Ker}(\Phi - \lambda I) = \left\lbrace X \in \mathcal{B}(\mathcal{H}) |\, U_i X U_i^{\dagger} = \lambda X, \quad \forall i \right\rbrace.
\end{equation} 
This reveals a constructive way how to obtain the attractors of the superoperator $\Phi$. The asymptotic state $\rho_\infty(n)$ of the system with the initial state $\rho(0)$ for $n\gg 1$ can be written in the form
\begin{equation}\label{asymprotic_state}
	\rho_{\infty}(n) = \sum_{\lambda\in\sigma_{\abs{1}}, i} \lambda^n \Tr[X_{\lambda,i}^{\dagger} \rho(0)]X_{\lambda,i},
\end{equation}
where the index $i$ represents different orthonormal attractors corresponding to the same eigenvalue $\lambda$.
It is important to notice that the final asymptotic state does not depend on the probability distribution $p_i$ at all. However, the speed of convergence towards this state might. Another thing which can be seen from the formula (\ref{asymprotic_state}) is that the asymptotic state does not necessarily need to be stationary, it can be periodic or even non-periodic. 

In case of discrete-time quantum walks the random unitary operators can be written in the form $U_K = S_K C$, so the condition from (\ref{konstrukce_attraktoru}) on attractors corresponding to the eigenvalue $\lambda$ becomes
\begin{equation}\label{attractor_condition}
	S_K C X C^{\dagger} S_K^{\dagger} = \lambda X, \quad \forall K\subset E.
\end{equation}
Taking into account the special case of the shift operator (\ref{shift_operator_K}) and considering the case of an empty configuration, i.e. no edges present, one obtains
\begin{equation}\label{coin_condition}
	RC X (RC)^{\dagger} = \lambda X.
\end{equation}
Here $R$ is a global reflection operator
\begin{align}\label{global_reflection}
	R = \sum_{i}\mathcal{R} \dyad{i,R} + \mathcal{R} \dyad{i,L}.
\end{align}
The equation (\ref{coin_condition}) is called the coin condition. It has a useful block structure. Indeed, both $R$ and $C$ have a block structure in the context of vertex subspaces. After splitting the attractor $X$ into vertex blocks 
\begin{equation}\label{vertex_blocks}
	X^{v_1}_{v_2} \equiv \bra{v_1}X\ket{v_2}, \quad \forall v_1,v_2 \in \lbrace 0,1,...,N-1 \rbrace,
\end{equation}
the coin condition for the vertices reads
\begin{equation}\label{coin_condition_block_wise}
	\mathcal{R} C X^{v_1}_{v_2} C^{\dagger}\mathcal{R}^{\dagger} = \lambda X^{v_1}_{v_2}.
\end{equation}
This can be equivalently reformulated into a simple eigenvector equation of the form
\begin{equation}\label{coin_condition_tensor}
	\left[ \mathcal{R} C \otimes \left( \mathcal{R}C\right) ^{*}\right]  x^{v_1}_{v_2}= \lambda x^{v_1}_{v_2},
\end{equation}
where $\braket{c,d}{x^{v_1}_{v_2}} = \bra{c} X^{v_1}_{v_2}\ket{d} $ with $ \ket{c},\ket{d} $ being the coin states on vertices $v_1, v_2$ respectively.
Through these conditions one obtains the possible blocks of the attractor.

The way to build an attractor from these blocks is provided by the shift conditions. Rewriting the equation (\ref{attractor_condition}) as
\begin{equation}
	C X C^{\dagger} = \lambda S_K^{\dagger} X S_K, \quad\forall K\subset E,
\end{equation}
it becomes obvious, that since the left side is the same for all configurations, the right one must be also. From that directly follow the shift conditions
\begin{equation}\label{shift_conditions}
	S_K^{\dagger} X S_K = S_L^{\dagger} X S_L , \quad K,L\subset E.
\end{equation}

Even though this split of the original condition (\ref{attractor_condition}) provides a more constructive way of looking for attractors, it is still in most cases a non-trivial problem. There is however a certain special subset of attractors for which the problem simplifies significantly. First, one needs to find the common eigenstates of all the possible unitary operators, i.e.
\begin{equation}\label{common_eigenstates}
	U_K \ket{\phi_{\alpha}} = \alpha \ket{\phi_{\alpha}}, \quad \forall K\subset E, \alpha \in \mathbb{C}.
\end{equation}
Then an arbitrary linear combination of the form
\begin{equation}\label{p-atraktor}
	Y_{\lambda} = \sum_{\lambda=\alpha \beta^{*}} A_{\beta,j}^{\alpha,i}\dyadB{\phi_{\alpha,i}}{\phi_{\beta,j}},
\end{equation}
is an attractor with eigenvalue $ \lambda=\alpha \beta^{*} $. In the formula above $A_{\beta,j}^{\alpha,i}\in \mathbb{C}$ are the linear coefficients and indices $i,j$ denote different common eigenstates related to the same eigenvalue. Attractors of this kind are called the p-attractors and they are easier to find, due to the reduced dimensionality of the problem. 

For common eigenstates one can actually obtain the coin condition of the form
\begin{equation}\label{coin_common_eigenstates}
	R C \ket{\phi_{\alpha}} = \alpha \ket{\phi_{\alpha}},
\end{equation}
and also the corresponding shift conditions
\begin{equation}\label{shift_common_eigenstates}
	S^{\dagger}_K \ket{\phi_{\alpha}} = S^{\dagger}_L \ket{\phi_{\alpha}}, \quad \forall K,L\subset E.
\end{equation}
in a very similar way as it has been done earlier for the general attractors. 

\section{Two-particle percolated quantum walks}
\label{sec:2pqw}

For the quantum walk of two particles on the same graph the Hilbert space is defined as
\begin{equation}
	\mathcal{H}^{(2)}=\mathcal{H}\otimes \mathcal{H}= (\mathcal{H}_{p}\otimes \mathcal{H}_{c})\otimes(\mathcal{H}_{p}\otimes \mathcal{H}_{c}).
\end{equation}
The evolution operator without interaction is
\begin{equation}\label{evolution_two}
	U^{(2)} = U\otimes U,
\end{equation}
where $U$ is the one-particle evolution operator defined as (\ref{evolution operator}). 
The two-particle superoperator describing the percolated quantum walk is now given as
\begin{equation}\label{superoperator_2}
	\Phi(\rho) = \sum_{K\subset E} p_K \left( U_K\otimes U_K\right)  \rho \left( U_K^{\dagger}\otimes U_K^{\dagger}\right),
\end{equation}
where $\rho$ is the two-particle density matrix and $U_K$ are the one-particle evolution operators corresponding to the graph configurations $K$. The superoperator $\Phi$ has the form of the LOCC operation \cite{nielsen_quantum_2010,chitambar_everything_2014}, hence it can only decrease initial entanglement.

With the same procedure as for the single-particle quantum walk, one arrives at the condition for two-particle attractors $ X $ and their corresponding eigenvalues $ \lambda\in\mathbb{C} $ of the form
\begin{equation}
\label{condition_attr_two_part}
	(U_K\otimes U_K)X(U_K\otimes U_K)^{\dagger}=\lambda X,\quad \forall K\subset E.
\end{equation}
The equations can be again split into the coin and the shift condition. The one-particle local coin operator is chosen as $H$ and the local reflection operator as $\mathcal{R}=\sigma_x$ as previously. We provide the explicit form of these conditions in \ref{app:coin} and \ref{app:shift}. As we show in the following section, one can construct the two-particle attractors utilizing the single-particle attractors and the exchange symmetry of the evolution (\ref{evolution_two}). Nevertheless, there are also genuine two-particle attractors which are not separable. The coin and shift conditions are then crucial for the proof of completeness of the constructed set of attractors, which we provide in \ref{app:completeness}.

\section{Construction of the attractor space}
\label{sec:construction}

\subsection{Finding p-attractors}

Starting again with the p-attractors, the general common eigenvector is of the form
\begin{equation}\label{common_general}
	\ket{\psi}=\sum_{s,t=0}^{N-1}\sum_{c,d\in\left\lbrace L,R\right\rbrace }\psi_{s,c,t,d}\ket{s,c,t,d},
\end{equation}
where the amplitudes $ \psi_{s,c,t,d} $ have to satisfy the following shift conditions
\begin{align}
	s\neq t:& \quad \psi_{s,R,t,R}=\psi_{s-1,L,t,R}=\psi_{s,R,t-1,L}=\psi_{s-1,L,t-1,L}, \label{shift-p-attractors1}\\
	s=t :&\quad \psi_{s,R,s,R}=\psi_{s-1,L,s-1,L}, \nonumber\\
	&\quad \psi_{s-1,L,s,R}=\psi_{s,R,s-1,L}.\label{shift-p-attractors2}
\end{align}
It is obvious that if one takes two one-particle common eigenstates, then their tensor product is again a two-particle common eigenstate. It can be easily shown that the resulting common eigenstates then satisfy stricter shift conditions, namely condition (\ref{shift-p-attractors1}) holds even for the case $ s=t $. Clearly, any linear combination of these separable common eigenstates will satisfy these conditions as well. The situation is the same when constructing two-particle p-attractors from one-particle p-attractors. Hence, a certain subspace of the two-particle attractors, but not necessarily the whole attractor space, can always be constructed purely from one-particle dynamic. It will be shown that the same holds for non-p-attractors. The remaining problem is then whether there are any additional attractors which do not satisfy these stricter 'separable' shift conditions. That means attractors which cannot be constructed from one-particle attractors. 

In case of one particle on a line with the Hadamard coin as a local coin operator there are altogether two common eigenstates, one for each eigenvalue of $ \mathcal{R}H $ \cite{One_dim_line_general_coin}:
\begin{align}
	\lambda_+ =& \frac{1}{\sqrt{2}}(1+i):\quad \ket{\phi_+}=\frac{1}{\sqrt{N}}\sum_{s=0}^{N-1}(-i)^{s}\ket{s,+}, \nonumber\\
	\lambda_- = &\frac{1}{\sqrt{2}}(1-i):\quad \ket{\phi_-}=\frac{1}{\sqrt{N}}\sum_{s=0}^{N-1}(i)^{s}\ket{s,-}.
\end{align}
The coin states $\ket{\pm}$ are the corresponding eigenvectors of $ \mathcal{R}H $, namely
\begin{align}
\ket{+} = & \frac{1}{\sqrt{2}} (i\ket{L}+\ket{R}), \quad  
\ket{-} = \frac{1}{\sqrt{2}} (-i\ket{L}+\ket{R}).
\end{align}
From these one-walker common eigenstates four two-particle common eigenstates can be constructed
\begin{align}
	\lambda_{++} = i:&\quad \ket{\Phi_{++}}=\ket{\phi_+}\otimes\ket{\phi_+}=\frac{1}{N}\sum_{s,t=0}^{N-1}(-i)^{s+t}\ket{s,+,t,+}, \nonumber\\
	\lambda_{--} = -i:&\quad \ket{\Phi_{--}}=\ket{\phi_-}\otimes\ket{\phi_-}=\frac{1}{N}\sum_{s,t=0}^{N-1}(i)^{s+t}\ket{s,-,t,-},\nonumber\\
	\lambda_{+-} = 1:&\quad \ket{\Phi_{+-}}=\ket{\phi_+}\otimes\ket{\phi_-}=\frac{1}{N}\sum_{s,t=0}^{N-1}(-1)^s(i)^{s+t}\ket{s,+,t,-},\nonumber\\
	&\quad \ket{\Phi_{-+}}=\ket{\phi_-}\otimes\ket{\phi_+}=\frac{1}{N}\sum_{s,t=0}^{N-1}(-1)^t(i)^{s+t}\ket{s,-,t,+}.
\end{align}
However, for the eigenvalue $ \lambda_{+-} = 1 $ there is one more linearly independent (but not orthogonal) non-separable common eigenstate 
\begin{equation}\label{w}
	\ket{\Phi_w}=\frac{1}{\sqrt{2N}}\sum_{s=0}^{N-1}\left( \ket{s,+,s,-}+\ket{s,-,s,+}\right).
\end{equation}
Existence of this eigenstate can be easily seen from its shape in the $\lbrace \ket{L},\ket{R} \rbrace$ basis
\begin{equation}\label{w_RL}
	\ket{\Phi_w}=\frac{1}{\sqrt{2N}}\sum_{s=0}^{N-1}\left( \ket{s,L,s,L}+\ket{s,R,s,R}\right),
\end{equation}
and the shift conditions (\ref{shift-p-attractors1}) and (\ref{shift-p-attractors2}). The first part of (\ref{shift-p-attractors2}) is completely independent from the rest of the shift conditions, so it is possible to set all other amplitudes except for these diagonal ones to zero, which yields $\ket{\Phi_w}$. 
Orthogonalization with respect to the previous states yields 
\begin{equation}
    \ket{\Phi_{w'}} =  \sqrt{\frac{N}{N-1}}\left[ \ket{\Phi_{w}}-\frac{1}{\sqrt{2N}}(\ket{\Phi_{+-}}+\ket{\Phi_{-+}})\right].
\end{equation}
From these common eigenstates 25 orthonormal p-attractors can be constructed following the formula (\ref{p-atraktor}).

As for the circle, due to the periodic boundary condition one needs to find out which of the common eigenvectors are periodic. The periodicity condition for $ \ket{\Phi_{++}},\ket{\Phi_{+-}},\ket{\Phi_{-+}},\ket{\Phi_{--}} $ reduces to $ (i)^{N}=i^{0}=1 $. This means that all of these common eigenvectors can be present only for a circles of length $ N=4k, \- \forall k\in\mathbb{N} $. The remaining eigenvector $ \ket{\Phi_{w}} $ has position independent amplitudes, so it will be present for circles of all lengths. Hence, for circles of length $ N=4k, \- \forall k\in\mathbb{N} $ the situation is identical to the case of a line. For $ N\neq 4k $ there is only one common eigenstate $ \ket{\Phi_{w}} $ for the eigenvalue 1.

\subsection{Non-p-attractors}

For one particle, the only non-p-attractor is the identity $ I^{(1)} $ \cite{One_dim_line_general_coin}. However, for the two-particle quantum walk quite a significant amount of non-p-attractors appears. We show that all of the remaining non-p-attractors can be constructed using one-particle attractors. 

Tensor product of two one-particle attractors is a two-particle attractor. If at least one of the one-particle attractors is a non-p-attractor, then the resulting two-particle attractor is also a non-p-attractor. With this knowledge, one can construct the following 9 linearly independent non-p-attractors
\begin{align}\label{attractors_tensor1}
	\lambda_{1}=1:&\quad \left\lbrace I^{(1)}\otimes\dyad{\phi_{+}},\- I^{(1)}\otimes\dyad{\phi_{-}},\- \dyad{\phi_{+}}\otimes I^{(1)},\- \dyad{\phi_{-}}\otimes I^{(1)},\- I\right\rbrace , \nonumber\\
	\lambda_{2}=i:& \quad \left\lbrace I^{(1)}\otimes\ket{\phi_{-}}\bra{\phi_{+}}, \ket{\phi_{-}}\bra{\phi_{+}}\otimes I^{(1)}\right\rbrace , \nonumber\\
	\lambda_{3}=-i:& \quad \left\lbrace I^{(1)}\otimes\ket{\phi_{+}}\bra{\phi_{-}}, \ket{\phi_{+}}\bra{\phi_{-}}\otimes I^{(1)}\right\rbrace .
\end{align}

To get the rest of the non-p-attractors we introduce the SWAP operator
\begin{equation}\label{SWAP}
	W=\sum_{s,t=0}^{N-1}\ \sum_{c,d\in\left\lbrace L,R\right\rbrace } \dyadB{s,c,t,d}{t,d,s,c}.
\end{equation}
The SWAP operator is both unitary and hermitian. It can then be easily shown that if $ X $ is an attractor with the respective eigenvalue $ \lambda\in\mathbb{C} $, $ WX $ is also an attractor belonging to the eigenvalue $ \lambda $. Hence, another 9 non-p-attractors can be constructed by multiplying all of the attractors (\ref{attractors_tensor1}) with $ W $. It easy to check that these new attractors and the original ones form a linearly independent set. After normalizing the first 9 attractors the norm does not change by applying the $W$ operator, so the new 9 attractors will also be normalized. This means there are in total 4 non-p-attractors for $ \lambda_{2}=i $, 4 for  $ \lambda_{3}=-i $ and 10 for $ \lambda_{4}=1 $.

Concerning the walk on a circle, all of the one-particle p-attractors satisfy the periodicity condition only for $ N=4k,$ $ \forall k\in\mathbb{N}$. In that case there are all of the non-p-attractors constructed above. For $ N\neq4k,$ $ \forall k\in\mathbb{N}$ only two non-p-attractors remain - the global identity $ I $ and the SWAP operator $W$. 

In summary, together with the p-attractors there are 21 attractors for the eigenvalue 1, 10 for $ i $, 10 for $ -i $ and 2 for -1. So in total there are 43 attractors for two particles on the finite line of length $ N $ and also for the circles of lengths $ N = 4k,$ $ \forall k\in\mathbb{N}$. For the circles of lengths $ N\neq4k,$ $ \forall k\in\mathbb{N}$, there are in total only 3 attractors and that is $I,$ $W$ and $\dyad{\Phi_w}$. 

Compared to the one-particle case it is interesting to notice a few points. First, there is surprisingly a great number of non-p-attractors for the two-particle case in comparison to the one-particle case with only the identity present. Next is the fact that the total number of attractors is fixed for all lengths $N$, as it has been the case for one walker as well, however, the number is much greater. In the non-percolated case the two-particle walk in one dimension (line/circle) can easily be mapped to one-particle walk in two dimensions (lattice/torus). However, it turns out that with the presence of percolations the mapping is not so straightforward. If the percolations were independent for each particle, i. e. each particle would walk on an independent percolated walk, the mapping to the percolated two-dimensional lattice, or torus, would be possible. But walkers on the same line would result in only certain types of allowed percolations on the lattice. The non-equivalence can be seen from the results of \cite{Kollar2014}, where the special case of the Hadamard coin on the two-dimensional lattice has been considered as well. For the one-particle Hadamard walk on a two-dimensional lattice of size $N\times N$ it was shown in \cite{Kollar2014}, that there will be $4(N+1)^2$ p-attractors and the global identity. For periodic boundary conditions, i. e. torus of size $N\times N$, there is $4(N+1)^2$ p-attractors for even $N$ and $N^2$ p-attractors for odd $N$. In both cases the only non-p-attractor is again the global identity. So not only the number of attractors depends on $N$, but the only non-p-attractor is the identity. The fixed number of attractors and a number of non-trivial non-p-attractors is therefore clearly caused by the common percolated line.

We note that constructed attractors are linearly independent, however, they do not form an orthogonal set. This complicates the understanding of the asymptotic evolution since the formula (\ref{asymprotic_state}) requires the attractors to be orthonormal with respect to the Hilbert-Schmidt scalar product. For this reason the rest of the paper is focused only on specific examples where the orthogonalization can be either done analytically or is at least numerically feasible.

\section{Asymptotic states for circles of lengths $N\neq 4k$}
\label{sec:not4}

Let us now investigate the circles of lengths $N\neq 4k$ in more detail. In this case the number of attractors reduces to 3. In addition, all remaining attractors correspond to the same eigenvalue 1. Hence, the resulting asymptotic state is stationary, as can be seen from the general formula (\ref{asymprotic_state}). 

First, we determine the orthonormal basis of the attractor space, which consists of normalized global identity $\frac{1}{2N} I$, normalized SWAP operator
\begin{equation}
	\frac{1}{2N}\nonumber W = \frac{1}{2N} \sum_{x,y=0}^{N-1}\sum_{i,j=\{L,R\}} \ket{x,i,y,j}\bra{y,j,x,i},
\end{equation}
and the projector onto the common eigenstate $\ket{\Phi_w}$
\begin{equation}
	\frac{1}{2N} F \equiv \ket{\Phi_w}\bra{\Phi_w}.
\end{equation}
Considering the form of $\ket{\Phi_w}$ in the standard basis (\ref{w_RL}) we find that $F$ can be written as
\begin{equation}
	F = \sum_{x,y=0}^{N-1}\sum_{i,j=\{L,R\}} \ket{x,i,x,i}\bra{y,j,y,j} .
	\label{F:projector}
\end{equation}
By performing the Gram-Schmidt orthogonalization on the three attractors above we obtain the following basis
\begin{align}
	A_1 & =  \frac{1}{2N} I, \nonumber \\
	A_2 & =  \frac{1}{\sqrt{4N^2-1}}\left(W - A_1\right),\nonumber  \\
	A_3 & =  \sqrt{\frac{2N+1}{4N(N+1)(2N-1)}}\left(F - A_1 - \sqrt{\frac{2N-1}{2N+1}} A_2\right).
\end{align}

Let us now investigate some reachable asymptotic states of the two-particle percolated quantum walk, which are obtained by the projection of the initial state onto the attractor space (\ref{asymprotic_state}). We begin with the initial state of the two particles localized at 0 vertex with an arbitrary pure coin state, which we decompose into the basis of Bell states for simplification of later results, i.e. state of the form
\begin{equation}
\label{init:00}
	\ket{\psi_0} = \ket{0,0}\otimes(a\ket{\psi^+} + b\ket{\psi^-} + c\ket{\phi^+} + d\ket{\phi^-}),
\end{equation}
where the coefficients $a,b,c,d\in\mathbb{C}$ satisfy the normalization condition
\begin{equation}
\label{normalization_condition}
	|a|^2 + |b|^2+|c|^2+|d|^2=1.
\end{equation}
Evaluating the Hilbert-Schmidt scalar products of the initial density matrix and the orthonormalized attractors $A_j$ one finds the asymptotic state of the form
\begin{equation}
\rho_\infty = \frac{1}{2N} A_1 + \frac{2N-1-4N|b|^2}{2N\sqrt{4N^2-1}}A_2 + \frac{(2N+1)|c|^2 + |b|^2-1 }{\sqrt{N(N+1)(4N^2-1)}}A_3.
\end{equation}
It can be shown that the same asymptotic state can be obtained even for non-localized initial states or for initial states localized at a different vertex then 0. More specifically, one can extend the initial positions of the walkers to equal superposition of states where they are at the same vertex and keep the local coin state identical for all positions, i.e. initial states of the form
\begin{equation}
	\ket{\psi_0} = \left( {\cal N}\sum_{x}\ket{x,x}\right) \otimes(a\ket{\psi^+} + b\ket{\psi^-} + c\ket{\phi^+} + d\ket{\phi^-}),
\end{equation}
where  ${\cal N}$ is a normalization factor.

For further analysis it is more suitable to express the asymptotic state in terms of the identity $I$, SWAP operator $W$ and the operator $F$. By doing that one finds the following 
\begin{align}
\label{asymp:state:1}
\rho_\infty = & \ c_1 I + c_2 W + c_3 F,
\end{align}
where we have denoted
\begin{align}
    \nonumber c_1 & = \frac{N+|b|^2-|c|^2}{2N(2N^2+N-1)}, \\
    \nonumber c_2 & =  \frac{N-|c|^2-(2N+1)|b|^2}{2N(2N^2+N-1)}, \\
    c_3 & = \frac{{(2N+1)|c|^2+|b|^2-1}}{2N(2N^2+N-1)}.
\end{align}
We can write $\rho_\infty$ explicitly as
\begin{equation}
\label{asymp1}
   \rho_\infty = \sum_{x,y=0}^{N-1}\sum_{i,j=\{L,R\}} \left(c_1\ketbra{x,i,y,j}{x,i,y,j} + c_2\ketbra{x,i,y,j}{y,j,x,i} + c_3\ketbra{x,i,x,i}{y,j,y,j}\right).
\end{equation}
Now it is straightforward to analyze the partial transpose \cite{PeresPPT1996,HorodeckiPPT1996} of the asymptotic state. Indeed, from the expression of $F$ (\ref{F:projector}) it is clear that its partial transpose is the SWAP operator $W$. Hence, the partial transpose of the asymptotic density matrix reads
\begin{equation}
\rho_\infty^{\mathrm{PT}} =  c_1 I + c_3 W + c_2 F
\end{equation}
The eigenvalues of this operator can be determined by direct calculation. We find that there are only three distinct eigenvalues
\begin{eqnarray}
 \lambda_1 & = &  \frac{1-2|b|^2}{2N}, \quad  \lambda_2 =  \frac{1-2|c|^2}{2N(2N-1)}, \quad 
 \lambda_3 =  \frac{N-1+2|b|^2 + 2N |c|^2}{2N(2N-1)(N+1)},
\end{eqnarray}
with multiplicities 1, $N(2N-1)$ and $(N+1)(2N-1)$, respectively. While $\lambda_3$ is always positive, $\lambda_1$ or $\lambda_2$ can be negative if $|b|^2>1/2$ or $|c|^2>1/2$. In such a case, the asymptotic state is NPT  (i.e. has non-positive partial transpose) and hence entangled. If both $|b|^2<1/2$ and $|c|^2<1/2$ the state is PPT (i.e. it has positive partial transpose) and consequently may have only bound entanglement \cite{HorodeckiBoundentanglement}. 

We point out that the asymptotic state (\ref{asymp:state:1}) depends explicitly only on the initial amplitudes $b$ and $c$. Hence, any superposition of $\ket{\psi^+}$ and $\ket{\phi^-}$ converges to the same state given by
\begin{equation}
    \rho_\infty = \frac{1}{2(2N^2+N-1)} \left(I + W - \frac{1}{N} F \right) .
\end{equation}

Let us now turn to the probability distribution of the asymptotic state (\ref{asymp1}). The position distribution is given by
\begin{equation}
    w(x,y) = \sum_{i,j = \{L,R\}} \braketA{x,i,y,j}{\rho_\infty}{x,i,y,j} = 4 c_1 + 2(c_2+c_3)\delta_{x,y},
\end{equation}
which explicitly gives for non-diagonal terms
\begin{equation}
     w(x,y) = \frac{2N+2|b|^2-2|c|^2}{N(2N^2+N-1)}, \quad x\neq y,
\end{equation}
and for diagonal terms
\begin{equation}
    w(x,x) =
         \frac{3N-1+2(N-1)\left(|c|^2 - |b|^2\right)}{N(2N^2+N-1)} .
\end{equation}
If the parameters of the initial state satisfy
\begin{equation}
\label{asymp:max:mix}
    2(|b|^2 - |c|^2) = 1 - \frac{1}{N},
\end{equation}
we obtain a uniform position distribution $w(x,y) = 1/N^2$.  Note that the marginal (one-particle) distributions are always uniform.

Considering the reduced density matrix of the coins, we find that it reads
\begin{eqnarray}
\label{asymp1:redc}
    \nonumber \rho_C & = & \mathrm{Tr}_P \left(\rho_\infty\right) = c_1 N^2 I_C + c_2 N \sum_{i,j = \{L,R\}} \ketbra{i,j}{j,i} + c_3 N \sum_{i,j = \{L,R\}} \ketbra{i,i}{j,j} \\
   & = & \frac{1}{2(2N^2+N-1)} \begin{pmatrix}
r_1 & 0 & 0 & r_2 \\
0 & r_3 & r_4 & 0 \\
0 & r_4 & r_3 & 0 \\
r_2 & 0 & 0 & r_1
    \end{pmatrix} ,
\end{eqnarray}
where we have denoted
\begin{eqnarray}
    \nonumber r_1  = N^2+N-1 + N \left(|c|^2 - |b|^2\right)& , & \quad 
r_2  = (2N+1)|c|^2 + |b|^2-1, \\
 r_3 = N(N+|b|^2-|c|^2) & , & \quad 
    r_4 =  N-(2N+1)|b|^2-|c|^2 .    
\end{eqnarray}
One can show in a straightforward way that the reduced coin state (\ref{asymp1:redc}) is separable for all parameters of the initial state.


As another example we choose an initial state where the walkers start at arbitrary, but different positions $x$ and $y$. The initial state will then be
\begin{equation}
	\ket{\psi_0} = \ket{x,y}\otimes(a\ket{\psi^+} + b\ket{\psi^-} + c\ket{\phi^+} + d\ket{\phi^-}),\quad x\neq y,
\end{equation}
where the coefficients $a,b,c,d\in\mathbb{C}$ again satisfy the normalization condition (\ref{normalization_condition}). In this case the resulting asymptotic state does not depend on the choice of the initial coin state at all. Indeed, the asymptotic state will always be equal to
\begin{equation}
	\rho_\infty = \frac{1}{2N} A_1 - \frac{1}{2N\sqrt{4N^2-1}}A_2 - \frac{1 }{2N}\sqrt{\frac{N}{(N+1)(4N^2-1)}}A_3,
\end{equation}
which rewritten into $I$,$W$ and $F$ gives the formula
\begin{equation}
\label{asymp2}
	\rho_\infty = \frac{1}{4N(2N^2+N-1)}\left( (2N+1)I -  W -  F \right).
\end{equation}
The density matrix (\ref{asymp2}) corresponds to the previous result (\ref{asymp:state:1}) with the singlet initial state $(b=1)$. There are again more initial states which will result in the asymptotic state (\ref{asymp2}). These are for example all pure initial states where the coin state is the same for all positions and position-wise the walkers have zero probability of being on the same vertex, i.e. initial states of the form
\begin{equation}
	\ket{\psi_0} = \left( \sum_{x\neq y}f_{x,y}\ket{x,y}\right) \otimes(a\ket{\psi^+} + b\ket{\psi^-} + c\ket{\phi^+} + d\ket{\phi^-}),
\end{equation}

\section{Asymptotic states for circle or line of length 4}
\label{sec:length4}

Let us now turn to the case of length 4, where the results for the line and the circle coincide. In such a case, all 43 attractors constructed in Section~\ref{sec:construction} are present, and their orthogonalization has to be done numerically. Moreover, there are 4 different eigenvalues ($\pm 1, \pm i $), so the asymptotic state evolves with a period of 4 steps. We consider the initial state of the form (\ref{init:00}) and focus on the reduced states of coins and positions. 

\begin{figure}
    \centering
    \includegraphics[width=0.6\linewidth]{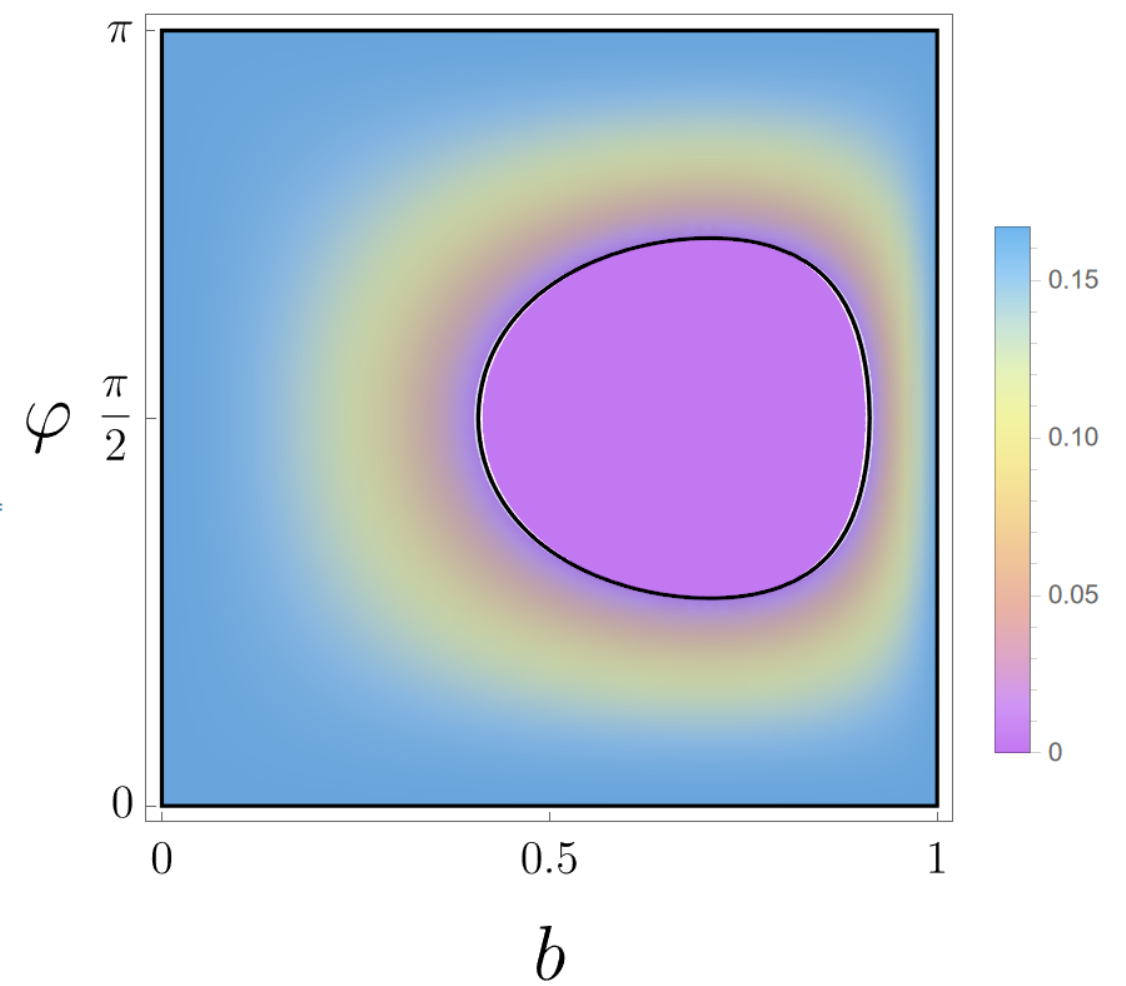}
    \caption{Concurrence of the asymptotic coin state (\ref{asymp:coin}) for the choice $q=0$ as a function of the parameters $b$ and $\varphi$ of the initial state. The analytical formula for the entanglement measure is given by (\ref{concurrence}). The black curve in the plot corresponds to the contour of the equality in (\ref{npt:cond}). }
    \label{fig:concurrence}
\end{figure}

We find that similarly to the examples studied in Section~\ref{sec:not4} the asymptotic coin state 
depends explicitly only on the amplitudes $b$ and $c$. If the initial state is a superposition of $\ket{\psi^+}$ and $\ket{\phi^-}$, i.e. $b=c=0$, the reduced coin state converges to a static asymptotic state
\begin{equation}
\label{asymp:ppt}
    \rho_C  = \frac{1}{4}\begin{pmatrix}
        1 & 0 & 0 & -\frac{1}{3} \\
        0 & 1 & \frac{1}{3} & 0 \\
        0 & \frac{1}{3} & 1 & 0 \\
        -\frac{1}{3} & 0 & 0 & 1
    \end{pmatrix} .
\end{equation}
One can easily check that the state is PPT and hence separable. On the other hand, for $\ket{\psi^-}$ and $\ket{\phi^+}$ the asymptotic coin states evolve periodically and are entangled. The explicit density matrices at times $4t+k$ are given by
\begin{eqnarray}
\nonumber  \rho_C^{(\psi^-)}(4t)   =  \rho_C^{(\phi^+)}(4t+2) & = & \frac{1}{8}\begin{pmatrix}
        1 & 0 & 0 & -\frac{1}{3} \\
        0 & 3 & -\frac{5}{3} & 0 \\
        0 & -\frac{5}{3} & 3 & 0 \\
        -\frac{1}{3} & 0 & 0 & 1
    \end{pmatrix} , \\
\nonumber \rho_C^{(\psi^-)}(4t + 1)  =  \rho_C^{(\phi^+)}(4t+3) & = & \frac{1}{4}\begin{pmatrix}
        1 & -\frac{1}{2} & \frac{1}{2} & \frac{1}{3} \\
        -\frac{1}{2} & 1 & -\frac{1}{3} & -\frac{1}{2} \\
        \frac{1}{2} & -\frac{1}{3} & 1 & \frac{1}{2} \\
        \frac{1}{3} & -\frac{1}{2} & \frac{1}{2} & 1
    \end{pmatrix} , \\
\nonumber \rho_C^{(\psi^-)}(4t + 2) = \rho_C^{(\phi^+)}(4t)  & = & \frac{1}{8}\begin{pmatrix}
        3 & 0 & 0 & \frac{5}{3} \\
        0 & 1 & \frac{1}{3} & 0 \\
        0 & \frac{1}{3} & 1 & 0 \\
        \frac{5}{3} & 0 & 0 & 3
    \end{pmatrix}, \\
\label{cycle:psim:phip}  \rho_C^{(\psi^-)}(4t + 3)  =  \rho_C^{(\phi^+)}(4t+1) & = & \frac{1}{4}\begin{pmatrix}
        1 & \frac{1}{2} & -\frac{1}{2} & \frac{1}{3} \\
        \frac{1}{2} & 1 & -\frac{1}{3} & \frac{1}{2} \\
        -\frac{1}{2} & -\frac{1}{3} & 1 & -\frac{1}{2} \\
        \frac{1}{3} & \frac{1}{2} & -\frac{1}{2} & 1
    \end{pmatrix} ,
\end{eqnarray}
so the asymptotic cycles of $\ket{\psi^-}$ and $\ket{\phi^+}$ are the same, only shifted by two steps. Concurrence \cite{Woottersconcurrence} of all four states (\ref{cycle:psim:phip}) equals $1/6$. In fact, concurrence of the asymptotic coin state is constant for any choice of the initial state. In order to evaluate it we choose the global phase such that $b>0$ and parameterize $c$ according to
\begin{equation}
    c = e^{i\varphi }\sqrt{1-b^2-q^2},\quad q^2 = |a|^2+|d|^2.
\end{equation}
The reduced coin states 
at times $4t+k$ are then given by
\begin{eqnarray}
  \rho_C(4t+k) & = &   \frac{1}{4}\begin{pmatrix}
        \rho_{11}(k) & \rho_{12}(k) & -\rho_{12}(k) & \rho_{14}(k) \\
        \overline{\rho_{12}}(k) & \rho_{22}(k) & \rho_{23}(k) & \overline{\rho_{12}}(k) \\
       - \overline{\rho_{12}}(k) & \rho_{23}(k) & \rho_{22}(k)  & -\overline{\rho_{12}}(k) \\
       \rho_{14}(k) & \rho_{12}(k) & -\rho_{12}(k) &  \rho_{11}(k)
    \end{pmatrix} ,
\label{asymp:coin}
\end{eqnarray}
where we have denoted
\begin{eqnarray}
\nonumber \rho_{11}(0) & = & \rho_{22}(2) = \frac{3-q^2}{2} - b^2 , \quad \rho_{22}(0) = \rho_{11}(2) = \frac{1+q^2}{2}+b^2 , \\
\nonumber \rho_{23}(0) & = & -\rho_{14}(2) = \frac{1+q^2}{6} - b^2, \quad \rho_{14}(0) =  - \rho_{23}(2) = \frac{5-7q^2}{6} - b^2, \\
\nonumber \rho_{12}(0) & = & -\overline{\rho_{12}}(2) = e^{i\varphi}b\sqrt{1-b^2-q^2} ,\\
\nonumber \rho_{11}(1) & = & \rho_{22}(3) = 1 - b\sqrt{1-b^2-q^2} \cos\varphi ,\\
\nonumber \rho_{22}(1) & = & \rho_{11}(3) = 1 + b\sqrt{1-b^2-q^2} \cos\varphi, \\
\nonumber \rho_{23}(1) & = & -\rho_{14}(3) = -\frac{1-2q^2}{3} - b\sqrt{1-b^2-q^2} \cos\varphi, \\
\nonumber \rho_{14}(1) & = & -\rho_{23}(3) = \frac{1-2q^2}{3} - b\sqrt{1-b^2-q^2} \cos\varphi, \\
\rho_{12}(1) & = & -\overline{\rho_{12}}(3)  = \frac{1-q^2}{2} - b^2 + i b\sqrt{1-b^2-q^2} \sin\varphi .
\end{eqnarray}
Calculating the partial transpose, we find that the states are NPT provided that the following inequality is satisfied
\begin{equation}
\label{npt:cond}
    b^2(1-b^2-q^2)\sin^2\varphi < \frac{1}{36}(5-26q^2+5q^4) . 
\end{equation}
Since the LHS is non-negative, if the overlap $q^2$ of the initial state with the subspace spanned by $\ket{\psi^+}$ and $\ket{\phi^-}$ is greater than $1/5$, the asymptotic coin state will be separable irrespective of $b$ and $\varphi$. For $q^2<1/5$ some of the initial entanglement can be preserved. Concurrence of the states (\ref{asymp:coin}) can be cast into the form
\begin{equation}
\label{concurrence}
    {\cal C} = \max\left(\frac{1}{12}\left(\sqrt{\alpha + \sqrt{\alpha^2-\beta^2}} - \sqrt{\alpha - \sqrt{\alpha^2-\beta^2}} - 4(1+q^2)\right),0\right), 
\end{equation}
where we have denoted
\begin{align}
\nonumber     \alpha & = 25 - 34q^2 + 13 q^4 - 72 b^2(1-b^2-q^2)\sin^2\varphi, \\
\beta & = 7 + 2q^2 - 5 q^4 .
\end{align}
For illustration, we display the concurrence (\ref{concurrence}) in Figure~\ref{fig:concurrence}. As the initial state we have considered a superposition of $\ket{\psi^-}$ and $\ket{\phi^+}$ only, corresponding to the choice $q=0$.

\begin{figure}
    \centering
    \includegraphics[width=0.47\textwidth]{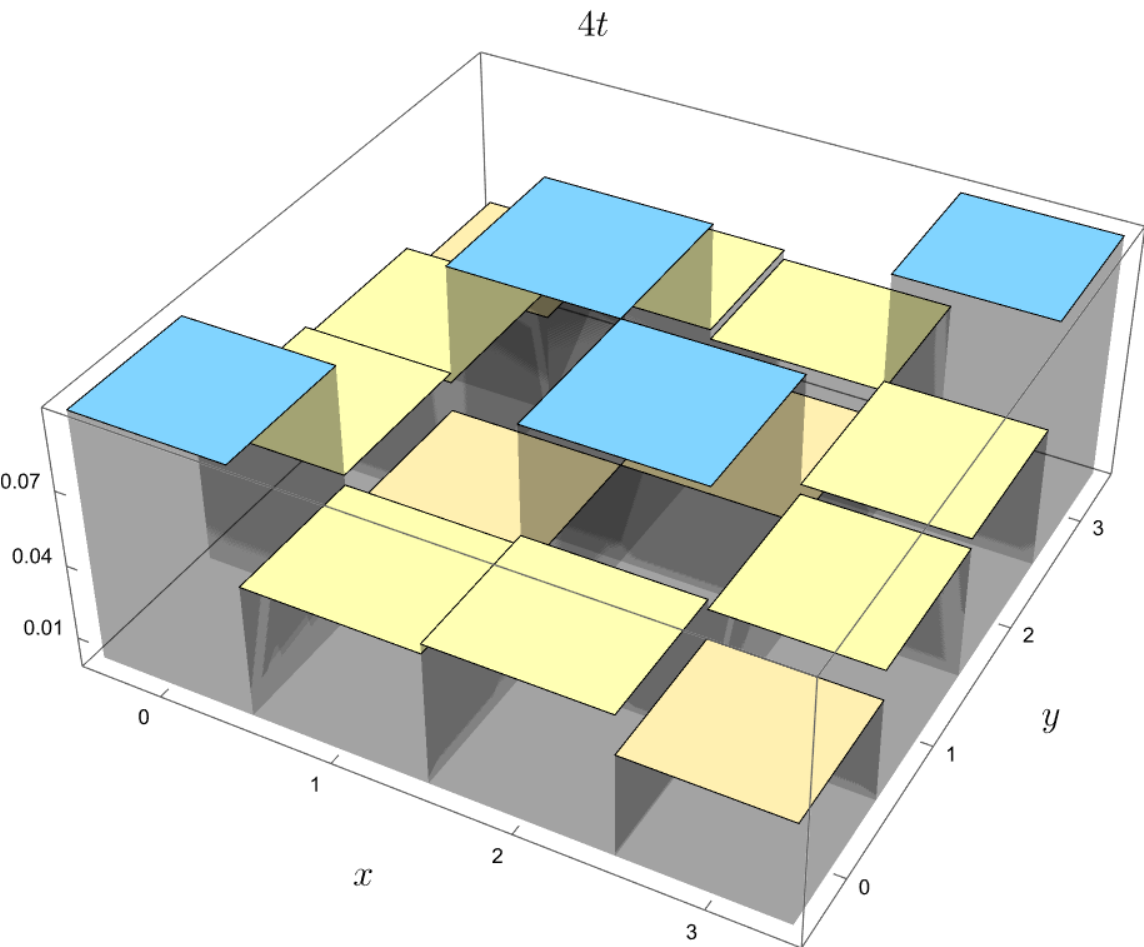}\hfill \includegraphics[width=0.47\textwidth]{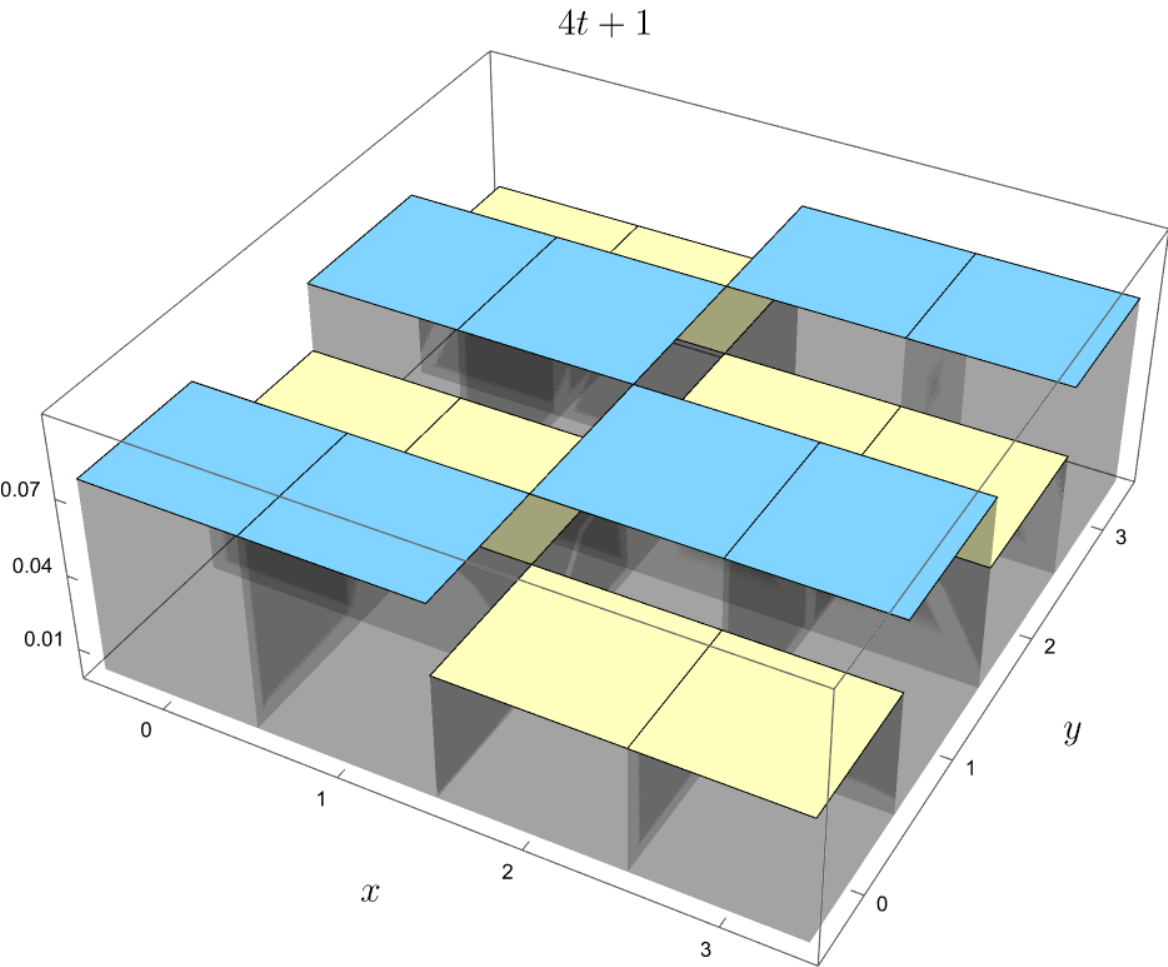}
    \includegraphics[width=0.47\textwidth]{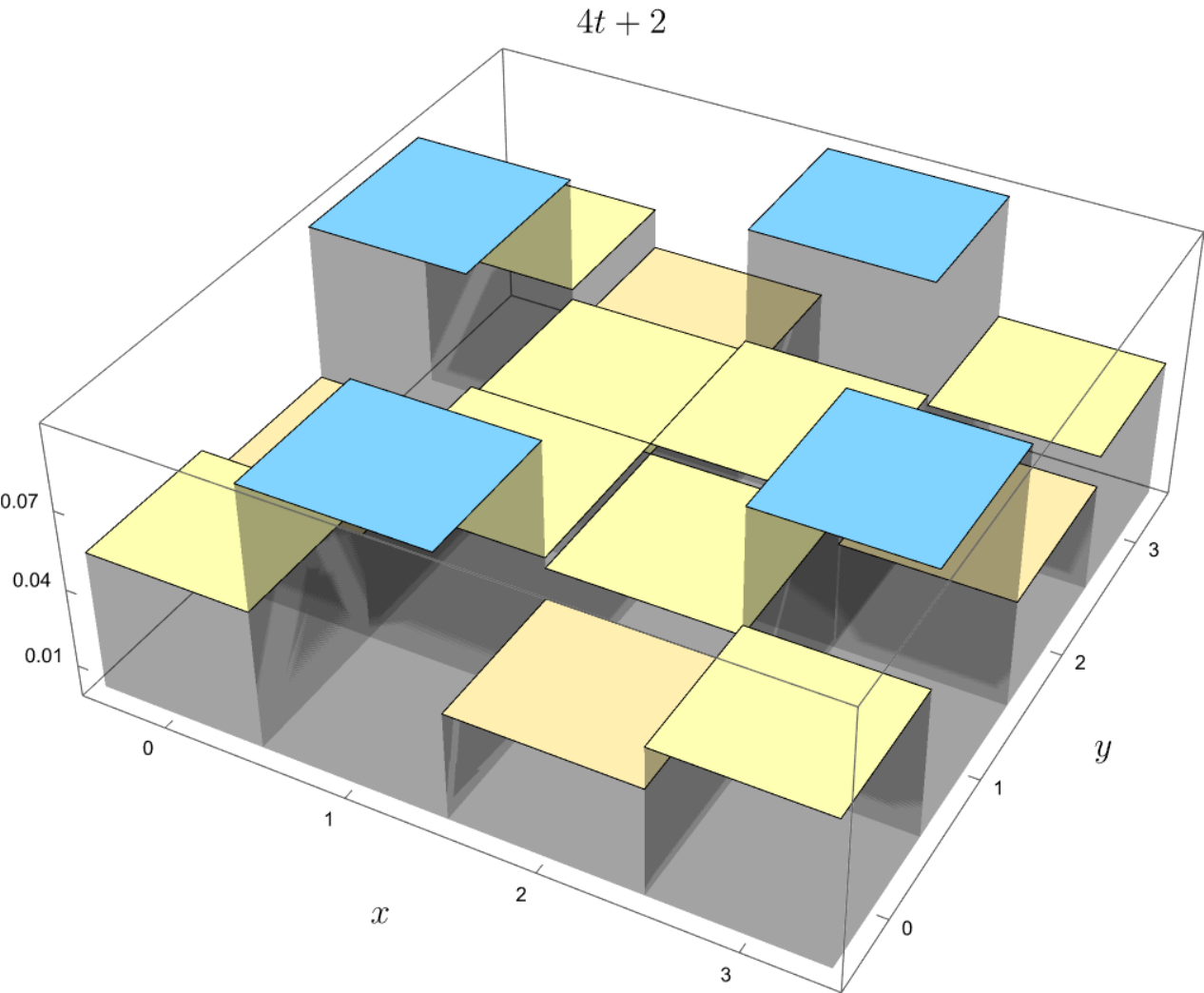}\hfill \includegraphics[width=0.47\textwidth]{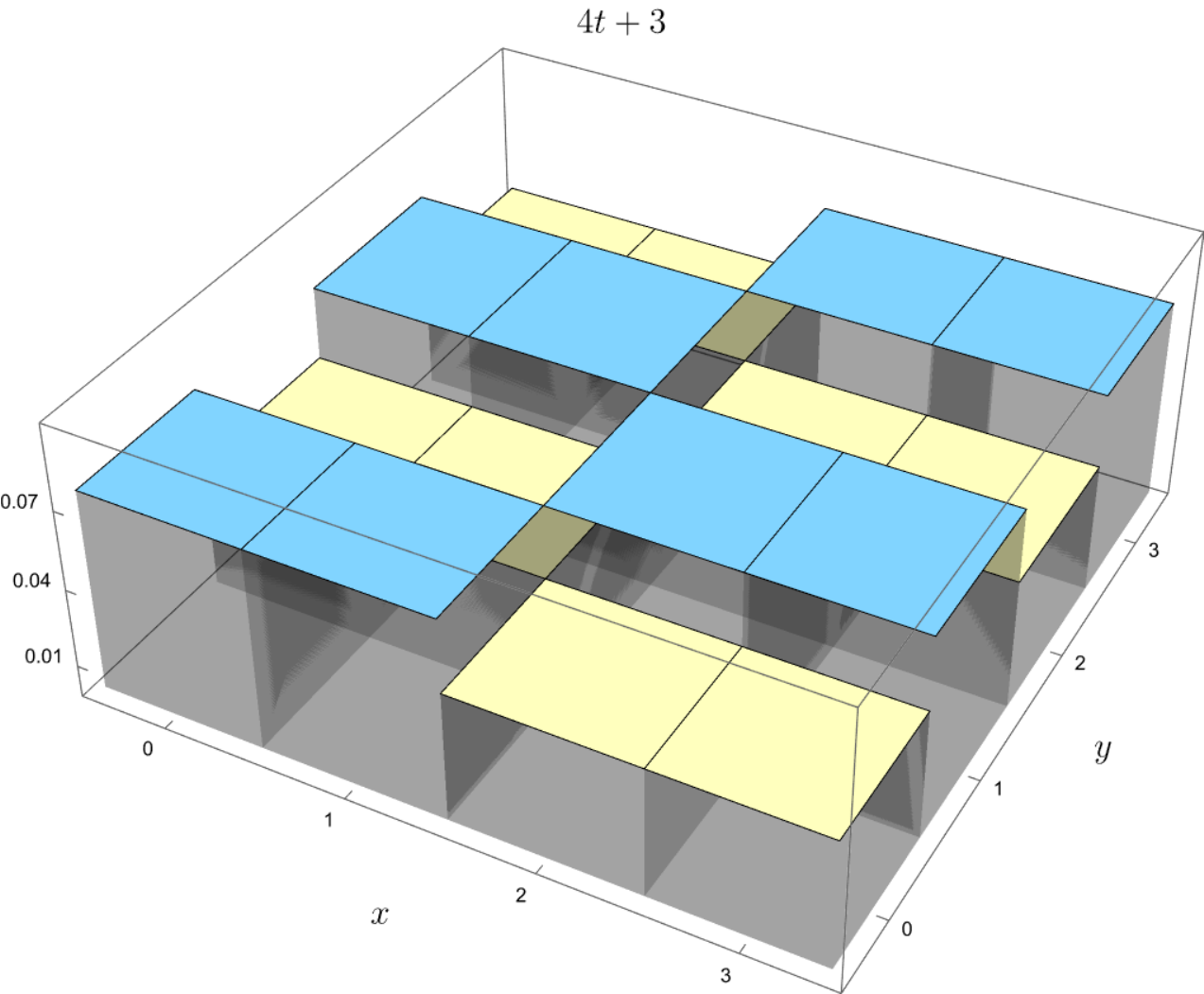}
    \caption{Asymptotic evolution of the probability distribution of the two-particle percolated quantum walk on a line or a circle of length 4 for the factorized initial state $\ket{LL}$. In this particular case, the distributions at times $4t+1$ and $4t+3$ are identical.}
    \label{fig:pd4}
\end{figure}

Let us now turn to the position distribution. After some algebra we find that it can be written in the form
\begin{eqnarray}
\nonumber  \big( w^{(4t)}(x,y) \big)  & = &  \frac{1}{3840}\begin{pmatrix}
        x_1 & x_2 & x_3 & x_4 \\
        x_3 & x_4 & x_1 & x_2 \\
        x_2 & x_1 & x_4 & x_3 \\
        x_4 & x_3 & x_2 & x_1 
    \end{pmatrix} , \\
\nonumber \big( w^{(4t+1)}(x,y) \big)  & = &   \frac{1}{3840}\begin{pmatrix}
        y_1 & y_2 & y_3 & y_4 \\
        y_3 & y_4 & y_1 & y_2 \\
        y_2 & y_1 & y_4 & y_3 \\
        y_4 & y_3 & y_2 & y_1 
    \end{pmatrix} , \\
\nonumber \big( w^{(4t+2)}(x,y) \big) &  = &   \frac{1}{3840}\begin{pmatrix}
        x_3 & x_4 & x_1 & x_2 \\
        x_1 & x_2 & x_3 & x_4 \\
        x_4 & x_3 & x_2 & x_1 \\
        x_2 & x_1 & x_4 & x_3 
    \end{pmatrix} , \\
\big( w^{(4t+3)}(x,y) \big) &  = &   \frac{1}{3840}\begin{pmatrix}
        y_3 & y_4 & y_1 & y_2 \\
        y_1 & y_2 & y_3 & y_4 \\
        y_4 & y_3 & y_2 & y_1 \\
        y_2 & y_1 & y_4 & y_3 
    \end{pmatrix} ,
\end{eqnarray}
where we have denoted
\begin{eqnarray}
\nonumber    x_1 & = &  294 + 3 |b|^2 + 49 |c|^2 + 24 |d|^2 + 52(\overline{c} d + c \overline{d}), \\
\nonumber x_2 & = & 306 - 3 |b|^2 - 49 |c|^2 - 24 |d|^2 - 52 (c \overline{d} + \overline{c} d ) , \\
 \nonumber x_3 & = & 3\big(58 + |b|^2 + 3 |c|^2 + 8 |d|^2 + 4 (c\overline{d} + \overline{c} d)\big) , \\
 x_4 & = & 3\big(62 - |b|^2 - 3 |c|^2 - 8 |d|^2 - 4(c \overline{d} + \overline{c} d ) \big)  ,
\end{eqnarray}
and 
\begin{eqnarray}
\nonumber y_1 & = & 234 + 3 |b|^2 + 29 |c|^2 + 24 |d|^2 -10 (b\overline{c} + \overline{b} c) - 20 (b \overline{d} + \overline{b} d ) + 32 (c\overline{d} + \overline{c} d), \\
\nonumber y_2 & = & 246 - 3 |b|^2 - 29 |c|^2 - 24 |d|^2 +10 (b\overline{c} + \overline{b} c) + 20 (b \overline{d} + \overline{b} d ) - 32 (c\overline{d} + \overline{c} d), \\
\nonumber y_3 & = & 234 + 3 |b|^2 + 29 |c|^2 + 24 |d|^2 +10 (b\overline{c} + \overline{b} c) + 20 (b \overline{d} + \overline{b} d ) + 32 (c\overline{d} + \overline{c} d), \\
\nonumber y_4 & = & 246 - 3 |b|^2 - 29 |c|^2 - 24 |d|^2 -10 (b\overline{c} + \overline{b} c) - 20 (b \overline{d} + \overline{b} d ) - 32 (c\overline{d} + \overline{c} d). \\
\end{eqnarray}
In contrast to the reduced coin state, the position probability distribution depends also on the amplitude $d$ of the initial state ($a$ was eliminated using the normalization condition (\ref{normalization_condition})). We see that the distribution has quite a nontrivial pattern, and in fact cannot become uniform. For illustration, we display in Figure~\ref{fig:pd4} the asymptotic cycle of the probability distribution for the factorized initial state $\ket{LL}$.

\section{Conclusions}
\label{sec:concl}

In the present paper we have investigated the asymptotic evolution of two-particle Hadamard walk on a dynamically percolated finite line and circle. Utilizing the known attractors for a single-particle case and the particle exchange symmetry, the attractors of the two-particle problem were constructed. Moreover, we have proven the completeness of our solution. The number of attractors increases dramatically in comparison to the one-particle case (from 5 to 43), resulting in greater sensitivity to the initial state and richer asymptotic dynamics. Nevertheless, the dimension of the attractor space remains independent of the length of the line $N$, which is in contrast to the single-particle walk on a 2D lattice, where it increases with $N$.

Two particular examples were studied in more detail. For circles of length $N\neq 4k$ the boundary conditions reduce the number of attractors significantly such that their orthogonalization can be performed by hand. Since all remaining attractors correspond to the same eigenvalue 1, the evolution converges to a steady state. For length $N=4$ all 43 attractors are present, nevertheless, it was possible to do the orthogonalization numerically and retrieve the asymptotic cycles of the reduced states of coins and the position distribution. We have utilized the decomposition of the initial coin state into the Bell basis. It is surprising that in both cases the reduced density matrix of coins depends only on the overlaps of the initial coin state with the Bell states $\ket{\psi^-}$ and $\ket{\phi^+}$. Hence, any superposition of $\ket{\psi^+}$ and $\ket{\phi^-}$ converges to the same reduced coin state. The asymmetry of the evolution for the Bell states $\ket{\psi^-}$ and $\ket{\phi^+}$ on one hand, and $\ket{\psi^+}$ and $\ket{\phi^-}$ on the other, deserves a more detailed investigation. 

We have also focused on the entanglement of the asymptotic state. For circles of length $N\neq 4k$ the overall steady state is entangled provided that the overlap of the initial state with one of the Bell states $\lvert\psi^-\rangle$ or $\ket{\phi^+}$ is greater than $1/2$. Nevertheless, the reduced coin state is always PPT and hence separable. However, for line or circle of length 4 some of the initial entanglement between coins can be preserved. The concurrence of the reduced coin state was evaluated explicitly, and it was shown that it converges and does not change anymore in the asymptotic cycle. 

As demonstrated by the examples of Sections~\ref{sec:not4} and \ref{sec:length4}, dynamical percolation results in correlations of quantum walkers positions, even if the initial state is fully uncorrelated. In contrast, for two non-interacting quantum walkers on an ideal line or a circle, this factorized unitary evolution cannot introduce additional correlations. In a similar spirit, classical two-particle random walk on a percolated line or a circle converges to a uniform distribution, i.e. there are no correlations between classical walkers. Emergence of position correlations in two-particle percolated quantum walk is thus an interplay of a quantum evolution and classical stochasticity. For quantum walkers the common dynamically changing graph manifests as a weak form of interaction. Utilizing the necessary and sufficient condition for non-zero quantum discord \cite{dakic_necessary_2010} one can show that the correlations in the asymptotic state, which result from this interaction, are not purely classical. However, evaluating quantum discord \cite{ollivier_quantum_2001,henderson_classical_2001} is a highly non-trivial task which goes beyond the scope of the present paper. It is also an interesting question how does the percolation compare to direct interaction of the two particles \cite{Schreiber2012,Silberberg2012}, which typically results in a bound state formation \cite{Stefanak2011,Molecular_binding_interaction_2012,qin:co-walking:2014}.

Concerning potential experimental realization of the two-particle percolated walk, photonic time-multiplexing seem to be the most suitable platform. Single-particle walk was implemented in \cite{Nitsche_2016}, where dynamical changes of the graph were realized by programmable fast-switching electro-optical modulators. Together with optimized photon pair sources \cite{Harder:13} and recent advances in photonic time-multiplexing \cite{nitsche_local_2020,Pegoraro_2023} we anticipate the implementation of two-particle walk on a percolated graph with 3 to 5 vertices to be feasible.

\ack

MP, M\v S and JN are grateful for financial support from RVO 14000 and ”Centre for Advanced Applied Sciences”, Registry No. CZ.02.1.01/0.0/0.0/16 019/0000778, supported by the Operational Programme Research, Development and Education, co-financed by the European Structural and Investment Funds. MP, M\v S and JN  acknowledge the financial support from Czech Grant Agency project number GA\v CR 23-07169S. MP is grateful for financial support from SGS22/181/OHK4/3T/14. TK acknowledges support from the National Research, Development and Innovation Office of Hungary, project No. TKP-2021-NVA-04.

\section*{References}

\bibliography{biblio}
\bibliographystyle{unsrt}

\appendix

\section{Coin condition}
\label{app:coin}

For two particles the coin conditions following from the attractor equation (\ref{condition_attr_two_part}) read
\begin{equation}\label{coin_condition_general}
	(RC\otimes RC)X(RC\otimes RC)^{\dagger}=\lambda X, \quad \lambda\in\mathbb{C}.
\end{equation}
As a  result of the block structure of the attractors this condition can be solved for individual attractor blocks $ \tensor*{X}{^{(s,t)}_{(u,v)}} $ as
\begin{equation}\label{coin_condition_general_block}
	(\mathcal{R}H\otimes \mathcal{R}H)\tensor*{X}{^{(s,t)}_{(u,v)}}(\mathcal{R}H\otimes \mathcal{R}H)^{\dagger}=\lambda \tensor*{X}{^{(s,t)}_{(u,v)}}, \quad \lambda\in\mathbb{C}.
\end{equation}
An important role in this case plays the tensor product structure of the coin condition. As has been mentioned previously, the operator
\begin{equation}
	\mathcal{\mathcal{R}}H = \frac{1}{\sqrt{2}}\left(\begin{matrix}
		1&-1\\
		1&1
	\end{matrix}\right),
\end{equation}
has two eigenvalues and their respective orthonormal eigenvectors are the following
\begin{align}
	\lambda_+ =& \frac{1}{\sqrt{2}}(1+i):\quad \ket{+}=\frac{1}{\sqrt{2}}\left(\begin{matrix}
		i\\
		1
	\end{matrix}\right)=\frac{1}{\sqrt{2}} (i\ket{L}+\ket{R}), \nonumber\\
	\lambda_- = &\frac{1}{\sqrt{2}}(1-i):\quad \ket{-}=\frac{1}{\sqrt{2}}\left(\begin{matrix}
		-i\\
		1
	\end{matrix}\right)=\frac{1}{\sqrt{2}} (-i\ket{L}+\ket{R}).
\end{align}
That means that the operator $ (\mathcal{R}H\otimes \mathcal{R}H) $ has eigenvalues $ \left\lbrace 1,i,-i\right\rbrace  $ and their respective eigenvectors can be directly constructed through the tensor product.
It follows that the eigenvalues $ \lambda $ of the operator (\ref{coin_condition_general}) can be obtained as $ \left\lbrace 1,i,-i,-1\right\rbrace  $  together with the corresponding eigenbasis of a general attractor block. However, a simpler basis can be found, in which the general attractor blocks of the aforementioned eigenvalues can be written as
\begin{align}\label{gen_block_1}
	\lambda_{1} =& 1: \tensor*{X}{^{(s,t)}_{(u,v)}} = \left( \begin{matrix}
		a & -b & -c & d \\
		-e & f & a-d-f &-b-c-e\\
		b+c+e & a-d-f & f & e \\
		d & c & b & a
	\end{matrix} \right),
\end{align}
\begin{align}\label{gen_block_i}
	\lambda_{2} =& i:\tensor*{X}{^{(s,t)}_{(u,v)}} = \left( \begin{matrix}
		-d & a & b & -ia-ib-d \\
		c & -ia-ic-d & -ib-ic-d &-a-b-c+2id\\
		-a-b-c+2id & ib+ic+d & ia+ic+d & c \\
		ia+ib+d & b & a & d
	\end{matrix} \right),
\end{align}
\begin{align}\label{gen_block_-i}
	\lambda_{3} =&-i:\tensor*{X}{^{(s,t)}_{(u,v)}} = \left( \begin{matrix}
		-d & a & b & ia+ib-d \\
		c & +ia+ic-d & +ib+ic-d &-a-b-c-2id\\
		-a-b-c-2id & -ib-ic+d & -ia-ic+d & c \\
		-ia-ib+d & b & a & d
	\end{matrix} \right),
\end{align}
\begin{align}\label{gen_block_-1}
	\lambda_{4} =&-1:\tensor*{X}{^{(s,t)}_{(u,v)}}= \left( \begin{matrix}
		a & -b & -b & -a \\
		-b & -a & -a & b\\
		-b & -a & -a & b \\
		-a & b & b & a
	\end{matrix} \right).
\end{align}
This basis will be used for the rest of the paper.

\section{Shift conditions}
\label{app:shift}

The shift conditions for two walkers can be expressed in the form
\begin{equation}\label{shift_two_particle}
	(S_L S_K^{\dagger}\otimes S_L S_K^{\dagger})X(S_K S_L^{\dagger}\otimes S_K S_L^{\dagger})=X \quad \forall L,K\subset E.
\end{equation}
The elements of the general attractor $ X $ will be denoted as \begin{equation}
	\tensor*{X}{^{s,C,t,D}_{u,E,v,F}} \equiv \bra{s,C,t,D}X\ket{u,E,v,F},
\end{equation} 
where $s,t,u,v\in \lbrace 0,1,...,N-1\rbrace$ and $C,D,E,F$ are the corresponding coin states. The whole attractor block for the first walker on vertices $s,u$ and the second walker on vertices $t,v$ will be denoted as $X^{(s,t)}_{(u,v)}$. Let us now write explicitly the shift conditions for these matrix elements. If none of the position indices are the same, the conditions are
\begin{flalign}\label{shift-general}
	\begin{split}
		\tensor*{X}{^{s,R,t,R}_{u,R,v,R}}&=\tensor*{X}{^{s-1,L,t,R}_{u,R,v,R}}=\tensor*{X}{^{s,R,t-1,L}_{u-1,L,v-1,L}}=\tensor*{X}{^{s,R,t-1,L}_{u,R,v,R}}=\tensor*{X}{^{s-1,L,t,R}_{u-1,L,v-1,L}}=\\&=\tensor*{X}{^{s,R,t,R}_{u-1,L,v,R}}=\tensor*{X}{^{s-1,L,t-1,L}_{u,R,v-1,L}}=\tensor*{X}{^{s,R,t,R}_{u,R,v-1,L}}=\tensor*{X}{^{s-1,L,t-1,L}_{u-1,L,v,R}}=\\&=\tensor*{X}{^{s-1,L,t-1,L}_{u,R,v,R}}=\tensor*{X}{^{s,R,t,R}_{u-1,L,v-1,L}}= \tensor*{X}{^{s-1,L,t,R}_{u-1,L,v,R}}=\tensor*{X}{^{s,R,t-1,L}_{u,R,v-1,L}}=\\&=\tensor*{X}{^{s-1,L,t,R}_{u,R,v-1,L}}=\tensor*{X}{^{s,R,t-1,L}_{u-1,L,v,R}}=\tensor*{X}{^{s-1,L,t-1,L}_{u-1,L,v-1,L}}.
	\end{split}
\end{flalign}

In case only the upper position indices are the same ($ s=t\neq u \neq v $) one arrives at
\begin{align}\label{shift-1}
	\begin{split}
		\tensor*{X}{^{s,R,s,R}_{u,R,v,R}}&=\tensor*{X}{^{s,R,s,R}_{u-1,L,v,R}}=\tensor*{X}{^{s-1,L,s-1,L}_{u,R,v-1,L}}=\tensor*{X}{^{s,R,s,R}_{u,R,v-1,L}}=\tensor*{X}{^{s-1,L,s-1,L}_{u-1,L,v,R}}=\\&=\tensor*{X}{^{s-1,L,s-1,L}_{u,R,v,R}}=\tensor*{X}{^{s,R,s,R}_{u-1,L,v-1,L}}=\tensor*{X}{^{s-1,L,s-1,L}_{u-1,L,v-1,L}},
	\end{split} \nonumber\\ 
	\begin{split}
		\tensor*{X}{^{s-1,L,s,R}_{u,R,v,R}}&=\tensor*{X}{^{s,R,s-1,L}_{u-1,L,v-1,L}}=\tensor*{X}{^{s,R,s-1,L}_{u,R,v,R}}=\tensor*{X}{^{s-1,L,s,R}_{u-1,L,v-1,L}}=\tensor*{X}{^{s-1,L,s,R}_{u-1,L,v,R}}=\\&=\tensor*{X}{^{s,R,s-1,L}_{u,R,v-1,L}}=\tensor*{X}{^{s-1,L,s,R}_{u,R,v-1,L}}=\tensor*{X}{^{s,R,s-1,L}_{u-1,L,v,R}}.
	\end{split}
\end{align}
The row of equations (\ref{shift-general}) splits similarly into 2 rows also for the cases when any other two indices are the same and the rest is different. 

For the case when there are two pairs of identical indices, for example $ s=t \neq u = v $ the shift conditions become
\begin{align}\label{shift-2}
	\begin{split}
		\tensor*{X}{^{s,R,s,R}_{u,R,u,R}}&=\tensor*{X}{^{s-1,L,s-1,L}_{u,R,u,R}}=\tensor*{X}{^{s,R,s,R}_{u-1,L,u-1,L}}=\tensor*{X}{^{s-1,L,s-1,L}_{u-1,L,u-1,L}},
	\end{split} \nonumber\\ 
	\begin{split}
		\tensor*{X}{^{s-1,L,s,R}_{u,R,u,R}}&=\tensor*{X}{^{s,R,s-1,L}_{u-1,L,u-1,L}}=\tensor*{X}{^{s,R,s-1,L}_{u,R,u,R}}=\tensor*{X}{^{s-1,L,s,R}_{u-1,L,u-1,L}}, 
	\end{split}\nonumber\\ 
	\begin{split}
		\tensor*{X}{^{s,R,s,R}_{u-1,L,u,R}}&=\tensor*{X}{^{s-1,L,s-1,L}_{u,R,u-1,L}}=\tensor*{X}{^{s,R,s,R}_{u,R,u-1,L}}=\tensor*{X}{^{s-1,L,s-1,L}_{u-1,L,u,R}},
	\end{split} \nonumber\\ 
	\begin{split}
		\tensor*{X}{^{s-1,L,s,R}_{u-1,L,u,R}}&=\tensor*{X}{^{s,R,s-1,L}_{u,R,u-1,L}}=\tensor*{X}{^{s-1,L,s,R}_{u,R,u-1,L}}=\tensor*{X}{^{s,R,s-1,L}_{u-1,L,u,R}}.
	\end{split}
\end{align}
Again, the row of equations (\ref{shift-general}) splits similarly into 4 parts also for the rest of the cases with two pairs of same indices.

In case there are three identical indices, the row of shift equations again splits into 4 rows, for example for $ s=t=u\neq v $ in the following way
\begin{align}\label{shift-3}
	\begin{split}
		\tensor*{X}{^{s,R,s,R}_{s,R,v,R}}&=\tensor*{X}{^{s-1,L,s-1,L}_{s-1,L,v,R}}=\tensor*{X}{^{s,R,s,R}_{s,R,v-1,L}}=\tensor*{X}{^{s-1,L,s-1,L}_{s-1,L,v-1,L}},
	\end{split} \nonumber\\ 
	\begin{split}
		\tensor*{X}{^{s-1,L,s,R}_{s,R,v,R}}&=\tensor*{X}{^{s,R,s-1,L}_{s-1,L,v-1,L}}=\tensor*{X}{^{s,R,s-1,L}_{s-1,L,v,R}}=\tensor*{X}{^{s-1,L,s,R}_{s,R,v-1,L}}, 
	\end{split}\nonumber\\ 
	\begin{split}
		\tensor*{X}{^{s,R,s,R}_{s-1,L,v,R}}&=\tensor*{X}{^{s-1,L,s-1,L}_{s,R,v-1,L}}=\tensor*{X}{^{s,R,s,R}_{s-1,L,v-1,L}}=\tensor*{X}{^{s-1,L,s-1,L}_{s,R,v-1,L}},
	\end{split} \nonumber\\ 
	\begin{split}
		\tensor*{X}{^{s,R,s-1,L}_{s,R,v,R}}&=\tensor*{X}{^{s,R,s-1,L}_{s,R,v-1,L}}=\tensor*{X}{^{s-1,L,s,R}_{s-1,L,v-1,L}}=\tensor*{X}{^{s-1,L,s,R}_{s-1,L,v,R}}.
	\end{split}
\end{align}

Finally for the diagonal terms, i.e. when $ s=t=u=v $, one gets
\begin{align}\label{shift-diag}
	\begin{split}
		\tensor*{X}{^{s,R,s,R}_{s,R,s,R}}=\tensor*{X}{^{s-1,L,s-1,L}_{s-1,L,s-1,L}},\quad
		\tensor*{X}{^{s-1,L,s,R}_{s,R,s,R}}=\tensor*{X}{^{s,R,s-1,L}_{s-1,L,s-1,L}} ,
	\end{split}\nonumber\\ 
	\begin{split}
		\tensor*{X}{^{s,R,s,R}_{s-1,L,s,R}}=\tensor*{X}{^{s-1,L,s-1,L}_{s,R,s-1,L}},\quad 
		\tensor*{X}{^{s-1,L,s,R}_{s-1,L,s,R}}=\tensor*{X}{^{s,R,s-1,L}_{s,R,s-1,L}},
	\end{split}\nonumber \\
	\begin{split}
		\tensor*{X}{^{s-1,L,s-1,L}_{s,R,s,R}}=\tensor*{X}{^{s,R,s,R}_{u-1,L,u-1,L}},\quad
		\tensor*{X}{^{s,R,s-1,L}_{s,R,s,R}}\tensor*{X}{^{s-1,L,s,R}_{s-1,L,s-1,L}} ,
	\end{split}\nonumber\\ 
	\begin{split}
		\tensor*{X}{^{s,R,s,R}_{s,R,s-1,L}}=\tensor*{X}{^{s-1,L,s-1,L}_{s-1,L,s,R}}, \quad 
		\tensor*{X}{^{s-1,L,s,R}_{s,R,s-1,L}}=\tensor*{X}{^{s,R,s-1,L}_{s-1,L,s,R}},
	\end{split}
\end{align}
so this time the row of equations splits completely into 8 parts.

The difference between the finite line and the circle is in the boundary conditions mentioned above, which affect the shift conditions. For the finite line all of the above mentioned shift conditions are valid for all $ s,t,u,v\in\left\lbrace1,2,...,N-1 \right\rbrace  $.
For the case of the circle all of the shift conditions hold for all $ s,t,u,v\in\left\lbrace0,1,2,...,N-1 \right\rbrace  $. This means that an attractor on the circle also has to satisfy the shift conditions for the line plus some additional ones for the cases when any of the position indices are 0. Consequently, any attractor for the circle must also be an attractor for the finite line.

\section{Proof of completeness}
\label{app:completeness}

This section will show that there are no more linearly independent attractors besides the ones that have been constructed in Section~\ref{sec:construction}. The idea is to start from the block $ \tensor*{X}{^{(0,0)}_{(0,0)}} $, which has to be fully determined. The next step is then to use the shift conditions to get all of its 15 neighboring blocks ($\tensor*{X}{^{(1,0)}_{(0,0)}}, \tensor*{X}{^{(0,1)}_{(0,0)}} ,\tensor*{X}{^{(0,0)}_{(1,0)}} ,\tensor*{X}{^{(0,0)}_{(0,1)}} ,\tensor*{X}{^{(1,1)}_{(0,0)}} , \tensor*{X}{^{(0,0)}_{(1,1)}}, \tensor*{X}{^{(1,0)}_{(1,0)}}, $ $\tensor*{X}{^{(0,1)}_{(0,1)}}, \tensor*{X}{^{(1,0)}_{(0,1)}}, $ $\tensor*{X}{^{(0,1)}_{(1,0)}}, \tensor*{X}{^{(1,1)}_{(1,0)}}, $ $\tensor*{X}{^{(1,1)}_{(0,1)}}, \tensor*{X}{^{(1,0)}_{(1,1)}}, $ $\tensor*{X}{^{(0,1)}_{(1,1)}}, \tensor*{X}{^{(1,1)}_{(1,1)}} $) while setting as little additional parameters of these neighboring blocks as possible. Some lower bounds on the dimensionality of attractor eigenspaces have been obtained from the previous construction (21 for the eigenvalue 1, 10 for $ \mathrm{i} $, 10 for $ -\mathrm{i} $ and 2 for -1). The goal is now to show that the dimensions can not be higher than that.

The block $ \tensor*{X}{^{(0,0)}_{(0,0)}} $ is a diagonal block. As has been shown in \ref{app:shift}, this is the block for which the most restricting shift conditions hold. All of the equalities which hold for the diagonal block $ \tensor*{X}{^{(s,s)}_{(s,s)}} $ also hold for all of the other types of blocks. For the other blocks one just gets some additional equalities depending on the type of combination of vertex indices. This means that after determining the first 16 blocks, the rest of the attractor is determined as well. Hence, it is sufficient to focus only on $ 	\tensor*{X}{^{(0,0)}_{(0,0)}} $ and its 15 nearest neighbors.

\subsection*{Cases $ \mathrm{i},\mathrm{-i},1$}

The proofs for the eigenvalues $i, -i$ and 1 follow the same idea, so only the case for the eigenvalue $i$ will be shown explicitly. The procedure for $-i$ is then completely the same up to some complex conjugations, see the structures of the general blocks (\ref{gen_block_i}) and (\ref{gen_block_-i}). For -1 there are more free parameters. The proof is based on the following observation. If we choose the parameters of the block $ \tensor*{X}{^{(00)}_{(00)}} $, then the shift conditions from \ref{app:shift} provide us with equalities
\begin{align}\label{shift-0000}
	\begin{split}
		\tensor*{X}{^{0,L,0,L}_{0,L,0,L}}=\tensor*{X}{^{1,R,1,R}_{1,R,1,R}}
	\end{split},\nonumber\\ 
	\begin{split}
		\tensor*{X}{^{0,R,0,L}_{0,L,0,L}}=\tensor*{X}{^{0,R,1,R}_{1,R,1,R}},\quad
		\tensor*{X}{^{0,L,0,R}_{0,L,0,L}}=\tensor*{X}{^{1,R,0,R}_{1,R,1,R}}
	\end{split},\nonumber\\ 
	\begin{split}
		\tensor*{X}{^{0,L,0,L}_{0,R,0,L}}=\tensor*{X}{^{1,R,1,R}_{0,R,1,R}},\quad
		\tensor*{X}{^{0,L,0,L}_{0,L,0,R}}=\tensor*{X}{^{1,R,1,R}_{1,R,0,R}}
	\end{split},\nonumber\\ 
	\begin{split}
		\tensor*{X}{^{0,R,0,R}_{0,L,0,L}}=\tensor*{X}{^{0,R,0,R}_{1,R,1,R}},\quad
		\tensor*{X}{^{0,L,0,L}_{0,R,0,R}}=\tensor*{X}{^{1,R,1,R}_{0,R,0,R}}
	\end{split},\nonumber\\ 
	\begin{split}
		\tensor*{X}{^{0,R,0,L}_{0,R,0,L}}=\tensor*{X}{^{0,R,1,R}_{0,R,1,R}},\quad
		\tensor*{X}{^{0,L,0,R}_{0,L,0,R}}=\tensor*{X}{^{1,R,0,R}_{1,R,0,R}}
	\end{split},\nonumber\\ 
	\begin{split}
		\tensor*{X}{^{0,R,0,L}_{0,L,0,R}}=\tensor*{X}{^{0,R,1,R}_{1,R,0,R}},\quad
		\tensor*{X}{^{0,L,0,R}_{0,R,0,L}}=\tensor*{X}{^{1,R,0,R}_{0,R,1,R}}
	\end{split},\nonumber\\ 
	\begin{split}
		\tensor*{X}{^{0,R,0,R}_{0,R,0,L}}=\tensor*{X}{^{0,R,0,R}_{0,R,1,R}},\quad
		\tensor*{X}{^{0,R,0,R}_{0,L,0,R}}=\tensor*{X}{^{0,R,0,R}_{1,R,0,R}}
	\end{split},\nonumber\\ 
	\begin{split}
		\tensor*{X}{^{0,R,0,L}_{0,R,0,R}}=\tensor*{X}{^{0,R,1,R}_{0,R,0,R}},\quad
		\tensor*{X}{^{0,L,0,R}_{0,R,0,R}}=\tensor*{X}{^{1,R,0,R}_{0,R,0,R}}
	\end{split}.
\end{align}
As can be seen, this determines the elements $ ^{RR}_{RR} $ of all the neighboring blocks. To proceed further, we fix the remaining parameters of one of the neighboring blocks, and utilize the shift conditions again, which determines another element of all the other blocks. Hence, in each iteration of the procedure we reduce the number of free parameters by one, until we finally determine the attractor.

Explicitly, in the case of the eigenvalue $i$ the general attractor block is given by 4 parameters, see (\ref{gen_block_i}). This shows which elements give information about which parameters
\begin{align}
	&^{RR}_{RR},^{LL}_{LL}... \pm d, \quad &^{RR}_{RL},^{LL}_{LR} ... a, \nonumber\\
	&^{LL}_{RL},^{RR}_{LR}...b, \quad &^{LR}_{LL},^{RL}_{RR} ... c, \nonumber\\
	&^{LR}_{RL},^{RL}_{LR}... \mp (\mathrm{i}b+\mathrm{i}c+d), \quad &^{LR}_{LR},^{RL}_{RL}... \pm (\mathrm{i}a+\mathrm{i}c+d), \nonumber\\
	&^{LL}_{RR},^{RR}_{LL}... \mp (\mathrm{i}a+\mathrm{i}b+d), \quad &^{LR}_{RR},^{RL}_{LL}... -a-b-c+2\mathrm{i}d.
\end{align}
Following (\ref{shift-0000}) the $ d $ parameter of all neighboring blocks of $ \tensor*{X}{^{(00)}_{(00)}} $ is determined. Next, we choose the rest of e.g. the block $ \tensor*{X}{^{(10)}_{(00)}} $, that means its $ a,b$ and $c $ parameters. Applying shift conditions concerning only the first 16 blocks then gives
\begin{align}\label{+}
	&\tensor*{X}{^{1,R,0,L}_{0,L,0,L}}=\tensor*{X}{^{0,L,1,R}_{1,R,1,R}},\quad
	\tensor*{X}{^{1,R,0,R}_{0,L,0,L}}=\tensor*{X}{^{0,L,0,R}_{1,R,1,R}},\quad \tensor*{X}{^{1,R,0,R}_{0,R,0,L}}=\tensor*{X}{^{0,L,0,R}_{0,R,1,R}},\nonumber\\
	&\tensor*{X}{^{1,R,0,L}_{0,R,0,R}}=\tensor*{X}{^{0,L,0,R}_{1,R,0,R}},\quad \tensor*{X}{^{1,R,0,L}_{0,R,0,L}}=\tensor*{X}{^{0,L,0,R}_{1,R,1,R}},\quad...
\end{align}
Writing out the rest of the conditions, one can see that the element $ ^{LR}_{RR} $ of the other blocks is determined as well. This means the value of $ (-a-b-c+2\mathrm{i}d) $ is known. To proceed further, we choose e.g. the rest of $ \tensor*{X}{^{(11)}_{(00)}} $, this means setting its parameters $ a$ and $b $. Writing down once again the attached shift conditions as in equations (\ref{+}), one will see that in this case the elements $ ^{LL}_{RR} $ of the remaining blocks are now fixed as well. That means the value of $ (\mathrm{i}a+\mathrm{i}b+d) $ is now known. Together with the previous information about the block parameters this results in the knowledge of the values of $ d,c$ and $ (a+b) $. Finally, by setting the parameter $ a $ of $ \tensor*{X}{^{(10)}_{(01)}} $ the whole block has now been obtained and it is possible to once again use the shift conditions. This time they give the elements $ ^{LR}_{RL} $ of the remaining blocks. This means the value of $ (\mathrm{i}b+\mathrm{i}c+d) $ is known which now determines all of the parameters of the remaining blocks. 

Overall, during this process $ 4+3+2+1=10 $ parameters of the first 16 blocks in total can be chosen, the remaining ones are then fully determined by the shift conditions. Hence, the dimension of this eigenspace is maximally 10. Since 10 linearly independent attractors have already been constructed, it is clear that the dimension of this subspace is truly equal to 10.

The same procedure gives the upper bound on the dimensions of the $-i$ subspace equal to 10, while for the 1 eigenspace we find the upper bound provided by 6+5+4+3+2+1 = 21. In both cases they coincide with the number of linearly independent attractors constructed in Section~\ref{sec:construction}.

\subsection*{Case $ \lambda_4 =-1$}

For the last case of eigenvalue -1 the procedure results in the upper bound 3 on the dimension of this subspace. However, only 2 attractors for this subspace have been found. Hence, we have to show that the upper bound can be lowered. In order to prove this we will use the shift conditions beyond the first 16 blocks.

The general attractor block (\ref{gen_block_-1}) has two parameters. From conditions (\ref{shift-0000}) all of the $ a $ parameters (indices note which attractor block do the parameters $ a $ and $ b $ belong to) can be obtained
\begin{align}
	\tensor*{a}{^{00}_{00}} &= \tensor*{a}{^{11}_{11}} =-\tensor*{a}{^{01}_{01}} =  -\tensor*{a}{^{10}_{10}} = -\tensor*{a}{^{11}_{00}} = -\tensor*{a}{^{00}_{11}} = - \tensor*{a}{^{10}_{01}} = -\tensor*{a}{^{01}_{10}},\nonumber\\
	\tensor*{b}{^{00}_{00}} &= \tensor*{a}{^{01}_{11}} =-\tensor*{a}{^{10}_{11}} = - \tensor*{a}{^{11}_{10}} = -\tensor*{a}{^{11}_{01}} = \tensor*{a}{^{00}_{01}} = - \tensor*{a}{^{00}_{10}} = \tensor*{a}{^{01}_{00}} = \tensor*{a}{^{10}_{00}}.
\end{align}
This means that not only all of the $ ^{RR}_{RR} $ elements are known, but also elements $ ^{LL}_{LL} $, $ ^{RR}_{LL} $, $ ^{LL}_{RR} $, $ ^{RL}_{RL} $, $ ^{LR}_{LR} $, $ ^{RL}_{LR} $, $ ^{LR}_{RL} $. This knowledge can now be used to get some of the $ b $ parameters. Taking what is known about $ \tensor*{X}{^{(10)}_{(00)}} $ and using only the shift condition concerning our first 16 blocks gives
\begin{equation}
	\tensor*{a}{^{10}_{00}} = \tensor*{b}{^{11}_{11}} =-\tensor*{b}{^{00}_{11}} = - \tensor*{b}{^{00}_{11}} = -\tensor*{b}{^{10}_{10}} = -\tensor*{b}{^{01}_{01}} =  -\tensor*{b}{^{01}_{10}} = -\tensor*{b}{^{10}_{01}} (= \tensor*{b}{^{00}_{00}}).
\end{equation}
Getting the rest of the $ b $ parameters is a bit more complicated, once one of them is determined, say for example $ \tensor*{b}{^{00}_{10}} $, the rest can also be easily obtained. To see this, one only needs to realize, that if parameter $ b $ is known, all of the elements $ ^{RR}_{RL} $, $ ^{RR}_{LR} $, $ ^{LR}_{RR} $, $ ^{RL}_{RR} $, $ ^{LL}_{RL} $, $ ^{LL}_{LR} $, $ ^{RL}_{LL} $, $ ^{LR}_{LL} $ are known as well. Using shift conditions on these elements then results in
\begin{equation}
	\tensor*{b}{^{00}_{10}} =-\tensor*{b}{^{01}_{00}} = -\tensor*{b}{^{10}_{00}} =-\tensor*{b}{^{00}_{01}} =  \tensor*{b}{^{11}_{10}} = \tensor*{b}{^{01}_{11}} =  \tensor*{b}{^{11}_{01}} = \tensor*{b}{^{10}_{11}}.
\end{equation}
However, to get $ \tensor*{b}{^{00}_{10}} $, it is necessary to reach for shift conditions beyond the first 16 blocks. Starting with $ \tensor*{a}{^{10}_{10}} $, which is already known, and using some of these shift conditions results in the following series of equations
\begin{equation}
	\tensor*{a}{^{10}_{10}} = -\tensor*{b}{^{10}_{20}} =\tensor*{a}{^{00}_{20}} =  \tensor*{b}{^{00}_{10}}.
\end{equation}
This proves that $ \tensor*{a}{^{00}_{00}} $ and $ \tensor*{b}{^{00}_{00}} $ determines all of the other parameters. Hence, this subspace of attractors truly is only two-dimensional.
\end{document}